\documentclass{aa}
\usepackage[varg]{txfonts}
\usepackage[english]{babel}
\usepackage{graphicx}
\usepackage{amssymb}
\usepackage{multirow}
\usepackage{multicol}
\usepackage{tabularx}
\usepackage{tabu}
\usepackage{booktabs,subcaption,amsfonts,dcolumn}
\usepackage{hvfloat}
\usepackage{commath}

\usepackage{amsmath}
\usepackage{latexsym}
\usepackage{natbib}
\usepackage{microtype}
\bibpunct{(}{)}{;}{a}{}{,}
\usepackage{longtable}
\usepackage{rotating}
\usepackage{supertabular}
\usepackage[usenames,dvipsnames]{xcolor}
\usepackage{float}
\usepackage{stfloats}
\usepackage{lipsum}

\usepackage{silence}
\usepackage[bookmarks, colorlinks, breaklinks]{hyperref}
\hypersetup{linkcolor=blue,citecolor=MidnightBlue,filecolor=black,urlcolor=MidnightBlue}

\def\threehun{{\sc The Three Hundred }}

\newcommand{\NEW}[1]{\textcolor{black}{#1}}
\newcommand{\IB}[1]{\textcolor{black}{#1}}

\usepackage{subfiles} 

\begin{document}

    \title{Spherical bias on the 3D reconstruction of the ICM  density profile in galaxy clusters } 
    
    \author{
    I. Veronesi\inst{1} \and
    I. Bartalucci\inst{1} \and 
    E. Rasia\inst{2,3} \and
    S. Molendi\inst{1} \and
    M. Balboni\inst{1,5} \and 
    S. De Grandi\inst{1,4} \and
    F. Gastaldello\inst{1} \and
    C. Grillo\inst{1,6} \and 
    S. Ghizzardi\inst{1} \and 
    L. Lovisari\inst{1} \and
    G. Riva\inst{1,6} \and
    M. Rossetti\inst{1} 
    }

    \institute{
        INAF - Istituto di Astrofisica Spaziale e Fisica Cosmica di Milano, Via A. Corti 12, 20133 Milano, Italy \and
        INAF – Osservatorio Astronomico di Trieste, via Tiepolo 11, I-34131 Trieste, Italy \and
        IFPU – Institute for Fundamental Physics of the Universe, via Beirut 2, 34151, Trieste, Italy \and  
        INAF - Osservatorio Astronomico di Brera, via E. Bianchi 46, 23807 Merate (LC), Italy \and
        DiSAT, Università degli Studi dell’Insubria, via Valleggio 11, I-22100 Como, Italy \and
        Dipartimento di Fisica, Universitá degli Studi di Milano, Via G. Celoria 16, 20133 Milano, Italy 
        }
        
    \date{} 

    \abstract{
    X-ray observations of galaxy clusters are routinely used to derive radial distributions of ICM thermdynamical properties such as density and temperature. However, observations allow us to access quantities projected on the celestial sphere only, so that an assumption on the three-dimensional distribution of the ICM is necessary. 
    Usually, spherical geometry is assumed.   
    }{
    The aim of this paper is to determine the bias due to this approximation on the reconstruction of ICM density radial profile of a clusters sample and on the intrinsic scatter of the density profiles distribution\NEW{, in particular when clusters’ substructures are not masked}. 
    }
    {We used simulated clusters for which we can access the three-dimensional ICM distribution; in particular, we considered a sample of 98 simulated cluster drawn from \threehun project. For each cluster we simulated 40 different observations by projecting the cluster along 40 different lines of sight.
    \NEW{We extracted the ICM density profile from each observation assuming the ICM to be spherical distributed. For each line of sight we then considered the mean density profile over the sample and compared it with the three-dimensional density profile given by the simulations. The spherical bias on the density profile is thus derived by considering the ratio between the observed and the input quantities.} 
    We also study the bias on the intrinsic scatter of the density profile distribution performing the same procedure. 
    }
    {We find a bias on the density profile, $b_n$, smaller than $10\%$ for $R\lesssim R_{500}$ while it increases up to $\approx 50\%$ for larger radii. 
    The bias on the intrinsic scatter profile, $b_s$, is higher, reaching a value of $\approx 100\%$ for $R\approx R_{500}$. 
    We find that the bias on both the analysed quantities strongly depends on the morphology composition of the objects in the sample: for clusters that do not show large scale substructures, both $b_n$ and $b_s$ are reduced by a factor 2, conversely for systems that do show large scale substructures
    both $b_n$ and $b_s$ increase significantly.}
    {}

    
    \maketitle

    \section{Introduction}
    Galaxy clusters play a crucial role in the understanding of both astrophysical processes and large scale structure evolution. They are the largest virialized objects generated from small density fluctuations in the primordial era and grown hierarchically under their own gravity influence. Indeed, they trace the Universe evolution and composition so that important cosmological knowledge can be derived by studying their properties \cite[e.g.][]{Voit_2005, Allen}. 
Moreover, many astrophysical processes take place within galaxy clusters so that the baryonic component properties derived from clusters studies are used for astrophysics and fundamental physics studies \citep[e.g.][]{Arnaud, Tozzi_2001}. 
In this context, X-ray observations play a major role as they can detect the emission associated to the IntraCluster Medium (ICM), the hot and rarefied plasma that lies among the galaxies. This component contributes $\sim 15\%$ to the total cluster matter and represents the main baryonic component, as stars and galaxies are only a few percent. 
In particular, through X-ray observations we are able to derive ICM density and temperature profiles. These are crucial quantities to derive clusters properties, since they provide the stating point for mass measures. In particular, they are used to derive the total cluster mass through the hydrostatic equilibrium equation. Moreover, the ICM density profile is used to obtain the gas mass which is a largely used proxy of the total mass of clusters \citep{arnaud2010,krav2006,Pratt} and to measure the clusters mass function, hence the mass density $\Omega_m$, which is a fundamental quantity for cosmological studies. Therefore, the reconstruction of the ICM density profile assumes a significant part in cosmological studies. 
Moreover, in the era of precision astronomy and cosmology \citep{Allen, Planck_tension, Salvati} it is very important to be aware of the contribution of each systematic error source that can affects any measure and to quantify them.

Every astronomical observation carries an intrinsic aspect that could introduce systematic errors. In fact, observations can be informative only about projected quantities, so an assumption on the underlying geometry is necessary to recover three-dimensional properties, such as density profiles.
In this context cluster shape assume a crucial role.

It is well known that the mass distribution within galaxy clusters is generally not spherical, even though determining its three-dimensional shape is still an issue. 
Several studies investigated the problem using multi-wavelength techniques, such as combination of gravitational lensing, X-ray and Sunyaev-Zel'dovich effect observations, both on individual clusters and cluster samples, and found a quite general triaxial morphology \citep[and references therein]{SerenoII,Sereno_single}. However, some clusters appear more spherical and present smoother density profiles thanks to virialization processes, while others can present density inhomogeneities that make the cluster shape more intricate.  

However, it is a standard practice to assume spherical geometry when deriving cluster gas properties from X-ray observations. 
In fact, this makes the gas distribution deprojection process easier to implement and more computationally efficient. 
Moreover, it is widely assumed that any possible geometrical bias on single systems are averaged out when considering clusters samples, thanks to the triaxial orientations that are assumed to be randomly distributed. 
The systematic errors on the gas distribution reconstruction due to spherical assumption has been investigated throughout the years, typically by comparing different deprojection models applied to different theoretical morphology (see e.g. \citealt{Binney_Strimpel, Piffaretti}). The most recent work in the context of X-ray and Sunyaev–Zel’dovich observations is by \cite{Buote_I, Buote_II} where they investigate the spherical averaging of galaxy clusters shaped using different ellipsoidal models. They quantify the orientation-average bias and scatter for many observable and find generally small mean biases with substantial scatter for different view-orientations. 
However, such studies mostly investigate the impact of elliptical shapes instead of spherical ones, considering clusters with smooth density profiles and leaving inhomogeneities presence aside.
In the present work we investigate the bias due to the spherical assumption on the reconstruction of the ICM density profile, considering the presence of inhomogeneities substructures. 
In fact, differently from the cited works based on theoretical geometrical models, here we consider clusters that are simulated in a cosmological context, such that their shapes are not theoretically defined following a geometrical model but are given by the cosmological framework and the gravitational interaction with the environment.
For each simulated cluster we emulate an X-ray observation on the sky plane, considering numerous different lines of sight. By using standard X-ray analysis procedures, we extract the ICM density profile from each projected cluster assuming spherical geometry and compare it with the true density profile given by the simulations.
In this way we are able to quantify the bias introduced by the spherical assumption.

This paper is organized as follows. 
In Sect.~\ref{sec:Dataset_maps} we describe the composition of the simulated cluster sample and the procedure to produce mock X-ray maps. In Sect.~\ref{sec:Analysis} we discuss the mock maps analysis procedure while in Sect.~\ref{sec:Results} we present the main results of the analysis. In Sect.~\ref{sec:Discussion} we discuss the results and in Sect.~\ref{sec:Conclusion} we sum up the biases that arise from the spherical approximation in the gas density profile. 

We adopted a flat $\Lambda$-cold dark matter cosmology with $\Omega_m(0) = 0.3, \ \Omega_\Lambda(0) = 0.7, \ H_0 = 70~\mathrm{km \ Mpc^{-1} \ s^{-1}} $. 
We notice that the cosmological density values are those of the simulated sample, which, however, is characterized by a slightly lower Hubble constant ($H_{0, \rm{sim}}= 67.77 \mathrm{km \ Mpc^{-1} \ s^{-1}}$): this difference does not affect the results of this paper.


    \section{Dataset and mock maps production} \label{sec:Dataset_maps}
\subsection{Dataset}
\begin{figure}[t]  
        \centering
        \includegraphics[width =0.49\textwidth]{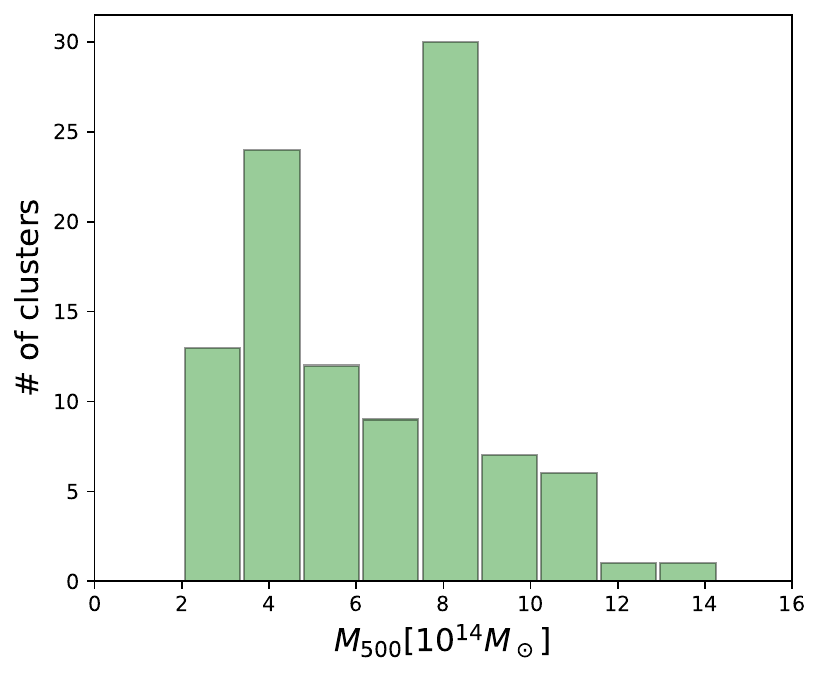} 
        \caption{Mass distribution of the sample of simulated clusters used in this work. The minimum and maximum values of the mass are  $2.0 \times 10^{14} M_\odot$ and $ 14.3 \times 10^{14} M_\odot$, respectively.}
        \label{fig:2_Sample_Mass}
    \end{figure}

\begin{figure*}[t]  
        \centering
        \includegraphics[width =\textwidth]{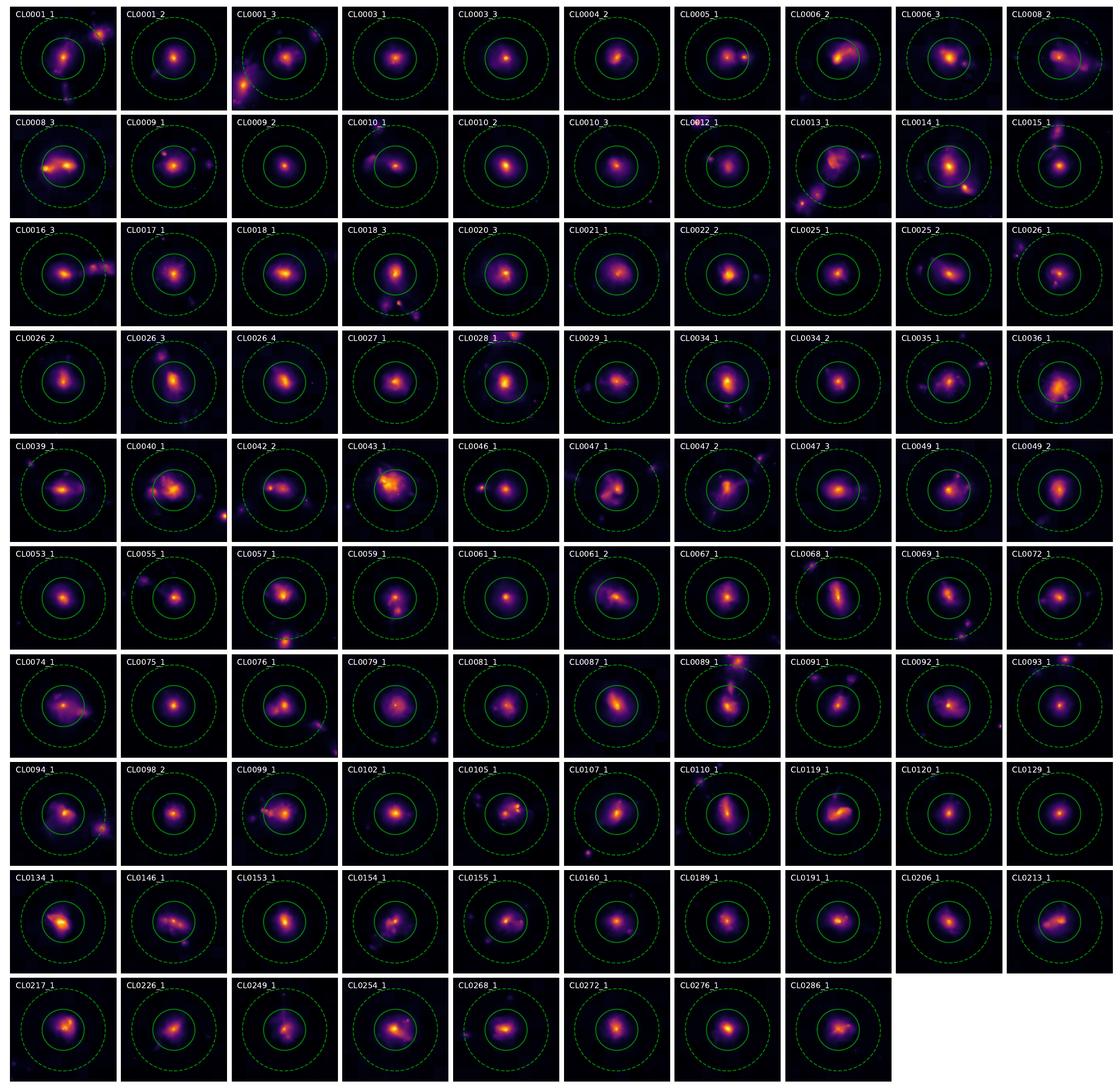} \par
        \caption{Image gallery of the 98 galaxy clusters in the simulated sample. Each image covers an area of $5R_{500} \times 5R_{500}$ and represent the spatial distribution of the ICM projected along one random direction in units of cm$^{-6}$Mpc; the solid circle indicates $R_{500}$ while the dashed one $2R_{500}$. In many clusters one or more substructures are present. The names of the simulated clusters are reported in each panel. The first number identifies the simulated region where the halo is extracted, while the second is the mass-rank index of the halo in that specific region. For example, CL0005\_1 is the most massive cluster found in the $5^{\rm th}$ region.}
        \label{fig:2_Sample_Gallery}
    \end{figure*}

    This work relies on simulated galaxy clusters drawn from \threehun project \citep{300Project}.  This project comprises 324 regions re-simulated with full-physics hydro-dynamical codes selected from the DM-only Multi-Dark Planck 2 Simulations \citep{MDPL2}. 
    The aim of this work is to statistically characterise the bias introduced in deriving the ICM radial density profile of clusters when a spherical geometry is assumed. 
    This is tested on a sample which needs to be as representative as possible of the underlying cluster population and able to reproduce the morphologies variety. 
    Therefore, we selected 98 clusters from \threehun catalogue with masses in the range $2.0 \times 10^{14} M_\odot < M_{500}\footnote{$M_{500} = \frac{4}{3} \pi R_{500}^3 500 \rho_c$, with $\rho_c$ critical density at cluster redshift.} < 14.3 \times 10^{14} M_\odot$  (Fig.~\ref{fig:2_Sample_Mass}). 
    \NEW{The sample was built to include about 20-25 objects in each of the following mass intervals: $[2-4]$, $[4-6]$, $[6-8]$, $[8-10] \ 10^{14} M_\odot$, in addition to 9 objects with $M_{500}>10^{15} M_{\odot}$. The selection was done by making sure that various morphologies were represented as shown in }   
    Fig.~\ref{fig:2_Sample_Gallery}: several clusters show substructures that appear as luminous regions distinguishable from the central halo while other have an almost spherical distribution.
    These features make the sample suitable to study the effects introduced by the spherical approximation, ordinarily used for X-ray analysis, and to quantify the related bias on the density profile reconstruction.
    All the clusters are fixed at redshift $z=0.3$, since we are not interested in studying any possible effect due to spatial resolution. 
    \NEW{The physical properties of each cluster are reported in Table~\ref{tab:A_1}.}
    
\subsection{Simulating mock X-ray maps }
    The starting point of this work is to create mock observations of the simulated clusters by projecting them on the sky plane and reproducing X-ray telescope effects, to obtain an Xray-like image. 
    
    \subsubsection{Projection on the sky plane}
\begin{figure}
            \centering
            \includegraphics[width = 0.24\textwidth]{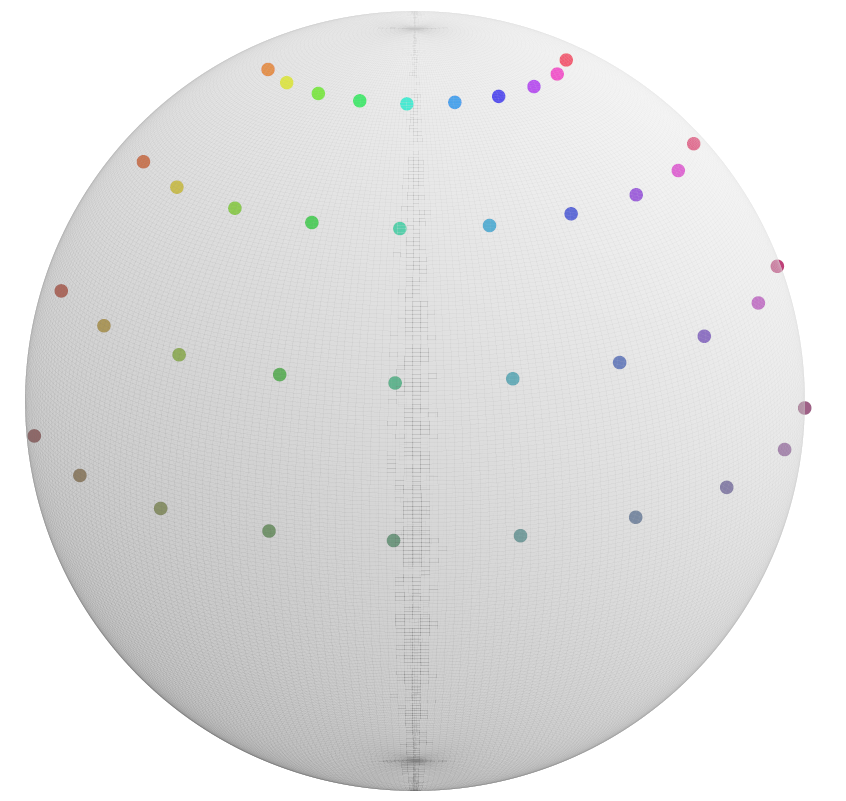}
            \hfill
            \includegraphics[width = 0.24\textwidth]{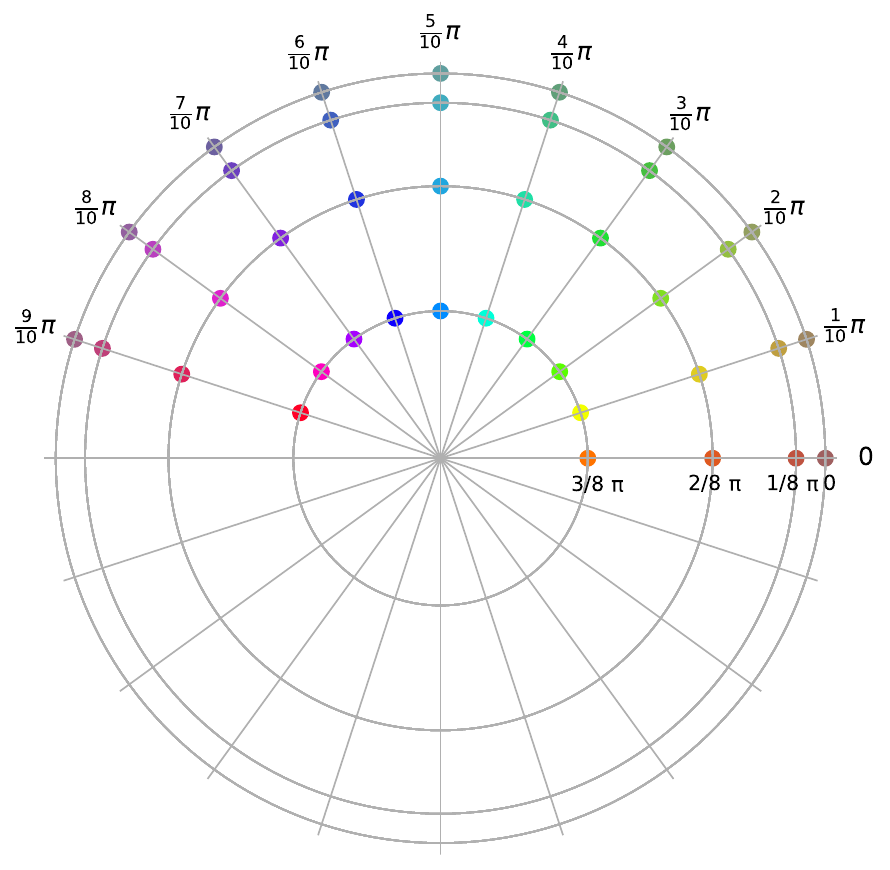}   
            \caption{Representation on the sphere (\textit{left}) and on the polar plane (\textit{right}) of the forty directions along which a three-dimensional simulated cluster is projected. 
            They are distributed at intervals of $\Delta\theta = 1/10 \ \pi$ and $\Delta\phi = 1/8 \ \pi$, with $\theta \in [0, 9/10 \ \pi]$ and $\phi \in [0, 3/8 \ \pi]$, where $\phi = 0$ identifies the equatorial plane. 
            Every direction will be identified as ($t, p$) where $t \in [0, 9]$ and $p \in [0, 3]$ will represent $\theta$ and $\phi$. 
            }
            \label{fig:2_angles}
        \end{figure}

        Since telescopes observe projected objects on the sky plane, we first projected each three-dimensional simulated cluster. 
        We created emission-measure maps from each simulated cluster with the program {\tt{Smac}} \citep[see][]{dolag.etal.2005,ansarifard.etal.20}. 
        The maps are centered at the position of the maximum of the density field which coincides with our definition of the theoretical center. 
        The field of view has a side equal to $6R_{500}$ divided in 1200 pixels leading to a resolution of 5\% $R_{500}$ per pixel. 
        This distance of $6R_{500}$ is also the length of the integration along the line of sight, reaching such high distance from the cluster center ($3R_{500} \approx 2R_{200}$ in the front and the back of the object) we are sure to include any possible large scale structure whose emission can be vision in the projected maps.
        Starting from a map, oriented as the $z$ axis of the simulation box and thus randomly oriented with respect to the cluster's major axis, we then created other maps obtained by rotating the object with equi-spaced angles as visually represented in Fig.~\ref{fig:2_angles}. 
        In this way we generated 40 emission measure maps for each simulated cluster. 
        The chosen 40 lines of sight allow to investigate the object from different directions, uniformly.
        
        The maps were produced summing over the contribution of all gas particles with temperature, $T_i$, larger than 0.3 keV and density, $\rho_i$, below the star-forming density threshold. This condition allow the study of the diffuse gas that emits in the considered X-ray bands. Notice that the same gas particle exclusion has been applied for all the quantities extracted from the simulated clusters such as the 3D gas density profiles. To create the maps we specifically sum the product ($m_i \times \rho_i$) of the selected particles once this contribution is  weighted by a spline kernel of width equal to the
        gas particle smoothing length. 

        The resulting maps are in units if a mass density squared,integrated over a volume, that is [$\mathrm{g^2 cm^{-6} \ Kpc^2 \ cm}$]. 
        The emission measure standard units are although [$\mathrm{cm^{-6}\ Mpc }$], so we needed to convert them, by taking into account the proton mass, the molecular weight $\mu$ and the electron-to-proton fraction $f$, since emission measure is referred to the electron emission. 
        We used the same gas parameters that characterize the simulated gas, that are $\mu = 0.59$ and $f = 1.08$.
        
    \subsubsection{Generating mock maps}\label{sec:2_mockMaps}
        Once the 40 emission measure maps has been generated for each cluster, we generated mock X-ray observations. 
        X-ray telescopes collect the incoming photons in the camera pixels, so they return counts maps. To create such maps from the emission measure ones we first converted them into surface brightness maps.       
        Emission measure ($EM$) and surface brightness ($SB$) are related through the cooling function $\Lambda(T,Z, z)$, which depends on the cluster temperature $T$, the cluster abundance $Z$ and the cluster redshift $z$:
            \begin{equation}\label{eq:2_SB_EM}
                SB = \Lambda(T,Z, z) \int n_e n_p dl = \Lambda(T,Z, z) EM(r)  .
            \end{equation}      
        
        In the soft X-ray band, the cooling function $\Lambda(T,Z, z)$ shows little dependence on both temperature and cluster abundance $Z$ \citep{Ettori, Bartalucci2017}. 
        To ensure this fact we computed the cooling function for different temperatures and abundances: with  typical cluster temperature values that can fluctuate by $20\%$ \citep[cfr. Fig. 11]{rossetti24} and typical cluster abundance values that can fluctuate by $25\%$ \citep[cfr. Fig.5 and Fig.11]{metal}, $\Lambda(T,Z)$ varies less than $3\%$.    
        Therefore the cooling function is computed using a constant temperature and constant abundance, within the $[0.5, 2.0] \ \mathrm{keV}$ energy band.
        The temperature for each cluster is given by the simulations, while the metal abundance is fixed to the average value $0.25 \mathrm{Z_\odot}$, as derived from \citep[cfr. Modified analysis of Table 1]{metal}. 
        The redshift is fixed for all clusters to $z = 0.3$. 
        The cooling function is computed via XSPEC \citep{XSPEC}, using the phabs (photoelectric absorbed) \footnote{https://heasarc.gsfc.nasa.gov/xanadu/xspec/manual/node259.html} APEC (Astrophysical Plasma Emission code)\footnote{https://heasarc.gsfc.nasa.gov/xanadu/xspec/manual/node134.html} model \citep{apec} and convolved with the telescope effective area. 
        For our scope, we considered the XMM-Newton PN camera.
        
        \renewcommand\arraystretch{1.2}
        \begin{table}[!b]
            \centering
            \caption{Parameters used in the phabs/APEC model for mock maps production and analysis. } \label{tab:2_Par}
            \begin{tabular}{ll}
                \toprule \toprule
                Flat $\Lambda$CDM cosmology  & $H_0=70 \ \mathrm{km/s/Mpc}$ \\
                                             & $\Omega_m=0.3$               \\
                                             & $\Omega_\Lambda=0.7$         \\
               \midrule
               Exposure time             & $30 \ \mathrm{ks}$                   \\
               Energy band               & $0.5-2.0 \ \mathrm{keV}$             \\
               Metallicity               & $0.25$  \citep{metal}                \\
               Temperature$^*$           & $2 \mathrm{keV}\lesssim T_{cl} \lesssim 10 \mathrm{keV}$ (Table~\ref{tab:A_1})                               \\ 
               $f = n_e/n_p$             & 1.08                                 \\
               Redshift                  & $0.3$ (from simulations)      \\
               Galactic column density$^{**}$   & $2.0\times 10^{-20} \ \mathrm{cm^{-2}}$ \citep{nh}  \\
               Background & $5.165 \times 10^{-3} \ \mathrm{cts \ arcmin^{-2} \ s^{-1}} \pm 3\%$  \\
               \bottomrule
            \end{tabular}
            \vspace{1ex}
            {\raggedright{\textbf{$^*$} The temperature is different for each cluster and it is given by the simulation. 
            
            \textbf{$^{**}$} The galactic column density is evaluated at high galactic latitudes.}\par}
        \end{table}

\begin{figure}[t]
            \centering
            \includegraphics[width =0.49\textwidth]{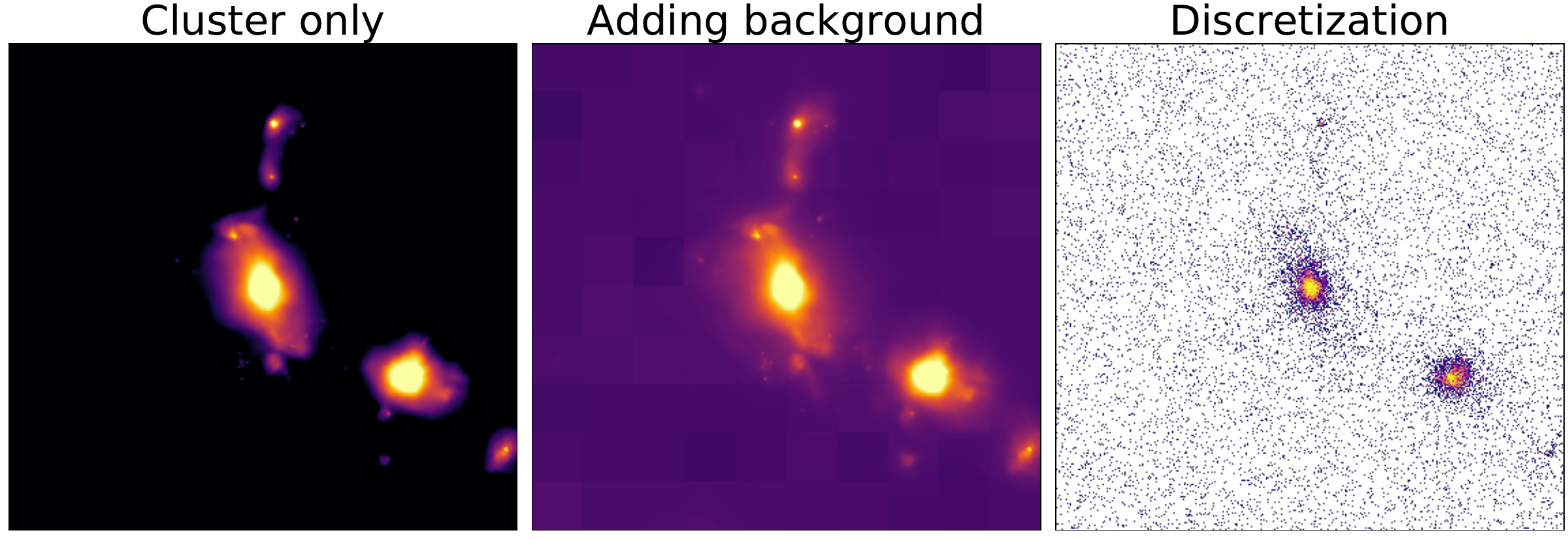}
            \caption{Mock map generation process.
            \textit{Left panel:} map of the cluster emission. 
            \textit{Central panel:} map of the cluster emission with the addition of the tiled background.
            \textit{Right panel:} map resulting from the discretization of the cluster+background map through a Poisson randomization. All maps are in units of counts.        }
            \label{fig:2_Mock_production}
        \end{figure}

        \renewcommand\arraystretch{1}
        Once we generated the surface brightness maps, we converted therm into Xray-like counts maps,
        following the same method described in \cite{Bartalucci}, that we briefly report here.
        Firstly, we converted the surface brightness into photons counts by multiplying the surface brightness maps by the pixels surface and by an observation time of $30\ \mathrm{ks}$, as a typical XMM-Newton exposure time. 
        For simplicity, we did not consider PSF and vignetting effects or the presence of malfunctioning pixels or gaps. For these reasons the exposure time is the same for all the map pixels.    
        We then added the sky background component: we considered a mean value $bkg_{mean} = 5.165 \times 10^{-3}\ \mathrm{cts \ arcmin^{-2} \ s^{-1}}$ measured by PN camera in the $[0.5 - 2] \mathrm{keV}$ band \citep{Bartalucci} and we introduced spatial variations by dividing the field of view in square tiles of $\sim 2.35 \ \mathrm{arcmin}$ size where the background value varies by the order of 3\% \citep{bkg5} around $bkg_{mean}$.
        Finally, we applied a Poisson randomization to each pixel of the map to emulate the discretized photon counts.
        All the used parameters are reported in Table~\ref{tab:2_Par}.
        The outcome of this procedure is represented in Fig.~\ref{fig:2_Mock_production}.
        
        In this way, we created 40 mock X-ray maps - one for each projection line - for each one of the 98 simulated clusters.
        
    \section{Data Analysis} \label{sec:Analysis}
\begin{figure*}[t]  
        \centering
        \includegraphics[trim=0 330 0 30,clip, width =\textwidth]{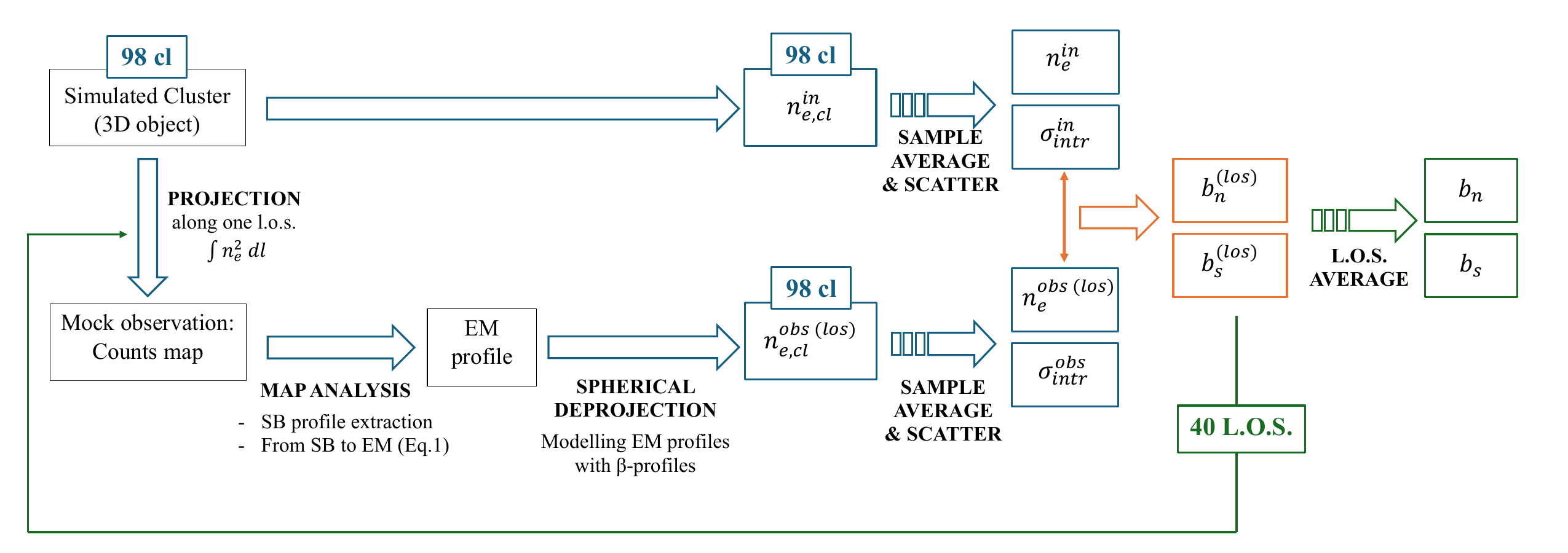} \par
        \caption{Schematic representation of the procedure described in Sec.~\ref{sec:2_mockMaps} and Sec.~\ref{sec:Analysis} to derive the gas density bias and the intrinsic scatter due to the spherical assumption. }
        \label{fig:3_procedure}
    \end{figure*}
    
The main purpose of the following analysis is to determine the bias on cluster gas density profile reconstruction due to the spherical symmetry approximation. In particular we evaluated the effect caused by the presence of inhomogeneity substructures. 
To reach this goal we used the mock X-ray maps obtained as described in Sect.~\ref{sec:2_mockMaps} to extract the gas density profiles following the same procedure used for real observations, where spherical symmetry is assumed. 
However, it is important to underline a key difference between the density profile extraction procedure followed in this work and the one used in real cluster analysis.
In fact, in the latter case if any substructure appears in the observation frame it is masked in such a way that its contribution to the cluster main emission will not be considered. 
Therefore it is interesting to study how the density profile is reconstructed even with the contribute of substructures, maintaining a spherical model. 
Furthermore, when we have to deal with real clusters observations, it can append that substructures are not distinguishable from the central core, because of resolutions issues of because they are located along the observer-core line of sight.
For these reasons it is important to evaluate the substructures impact on the gas density profile when it is reconstructed assuming a spherical shape. 

\subsection{Mock maps analysis}\label{sec:3_MockAnalysis}
    To extract the gas density profile from every mock X-ray map we used {\tt{pyproffit}}\footnote{https://pyproffit.readthedocs.io} \citep{pyproffit1,pyproffit2}, a python package largely used for X-ray cluster analysis, in which the spherical symmetry is assumed. 
        
    For each map {\tt{pyproffit}} first extracts the surface brightness profile, $SB$ (count rate per surface and time unit), by computing the mean surface brightness in concentric annular bins and dividing by the exposure time. We extract the $SB$ profile up to $3 R_{500}$ in annular bins of $5 \ \mathrm{arcsec}$ width. 

    The 
    $SB$ profile is then converted into emission measure profile, $EM$, through Eq.~\ref{eq:2_SB_EM}.
    The conversion factor $\Lambda(T,Z, z)$ is computed via XPSEC with the same procedure and parameters uesd for the mock maps production (see Sect.~\ref{sec:2_mockMaps} and Table~\ref{tab:2_Par}). 

    The obtained $EM$ profile is then deprojected to obtain the electron density profile. In fact, emission measure is defined as the projection along the line of sight of the gas density: 
        \begin{equation}
             EM(r) = \int n_e(R) n_p(R) dl = \frac{1}{f} \int n_e^2(R) dl \quad , 
        \end{equation}
    where $f$ is the electron-to-proton fraction.
    The $EM$ profile deprojection process to obtain the electron density profile is performed by {\tt{pyproffit}} assuming a spherical geometry, that is, the $EM$ profile is modeled as a combination of $\beta$-models: 
        \begin{equation}\label{eq:3_EM_beta}
             EM(r) =  \sum_i A_i \phi_i (r) = \sum_i A_i \Big[1+\Big(\frac{r}{r_{c, i}}\Big)^2\Big]^{-3\beta_i + \frac{1}{2}} \quad .
        \end{equation}
    In particular, we modelled the $EM$ profile by combining six $\beta$-models.
    Such a spherical geometry is widely used since the $\beta$-model functions $\phi_i$ can be analytically deprojected, returning the deprojected functions $\Phi_i (R) = \int \phi_i(r) dl$. 
    The combination of the deprojected functions $\Phi_i$ gives the electron density profile:
        \begin{equation} \label{eq:3_n}
            n_e (R) = \sum_i C_i \Phi_i(R) =  \sum_i C_i \Big[1+\Big(\frac{R}{r_{c, i}}\Big)^2\Big]^{-3\beta_i } \quad .
        \end{equation} 
    
    The $\beta$-models parameters ($\beta_i$, $r_{c,i}$, $A_i$) in Eq.~\ref{eq:3_EM_beta} are inferred by {\tt{pyproffit}} from the observed $EM$ profile by maximizing a poissonian likelihood. 
    The background is considered as a flat surface brightness profile, described by a single parameter: in particular we considered the radial range $[2.5, 3] R_{500}$ as background fitting region, where the background emission dominates over cluster emission.   
    The deprojection result is the electron density profile of the cluster (Eq.~\ref{eq:3_n}).
    The procedure is schematized in Fig.~\ref{fig:3_procedure}.

\subsection{Density Profiles Analysis}\label{sec:3_ProfAnalysis}
    We now study how the spherical symmetry assumption impacts on the clusters gas density profile reconstruction and in particular we can study how the presence of substructures influences the deprojection process. 

    The spherical assumption impact can be derived by comparing the extracted profiles with the input ones, that is, the density profiles given directly by simulations. 
    The input profiles are obtained by computing the density on spherical shells starting from the cluster center\footnote{
    We stress the fact that profiles computed using spherical shells can not be used to study the triaxial geometry of the cluster.}. 
    We divided the analysis studying firstly each cluster individually and then the sample as a whole. 

    \subsubsection{Single clusters analysis}\label{sec:3_ProfAnalysis_single}
        As a first step of the analysis we considered each cluster individually, so that we could compare the profiles extracted along each line ($n_{e,cl}^{obs (los)}$) of sight with the input density profile ($n_{e,cl}^{in}$) and study how well it is reconstructed by considering the ratio between each extracted profile and the input one. 
        
        This type of analysis shows how the spherical approximation works on specific clusters, allowing us to understand how substructures and their projected position (which depends on the line of sight) modify the reconstructed profile. 
        
    \subsubsection{Sample analysis} \label{sec:3_ProfAnalysis_sample}   
        As main core of the analysis we considered the entire sample observed only from one line of sight at a time: instead of considering every cluster from different lines of sight this approach would reproduce what a real observer can actually see.
        We can access at least 40 different realizations of the same sample (one for each projection line) and compare the results, giving us the opportunity to study how different projections impact on the density profile reconstruction.
        To perform this type of analysis we defined the sample global electron density profile $n_e$ and the related intrinsic scatter $\sigma_{intr}$.
        These quantities are defined as follow. 
        The global density profile is the logarithmic average profile of the 98 cluster profiles $n_{e,cl}$ in the sample; assuming that their distribution is log-normal then the logarithmic average is defined as the expected value of the logarithmic distribution of the 98 density profiles: 
        \begin{equation}
            n_{e} = \exp{\Big[\xi + \frac{\sigma^2}{2}\Big]},
        \end{equation}
        where $\xi$ and $\sigma$ are respectively the mean and the standard deviation of the normal distribution of $\log{n_{e,cl}}$.
        The intrinsic scatter $\sigma_{intr}$ represents the physical dispersion of the sample profiles around the global profile $n_e$ and it is related to the total dispersion $\sigma$ as: 
        \begin{equation}
            \sigma = \sqrt{\sigma_{intr}^2 + \sigma_{stat}^2},
        \end{equation}\label{eq:3_totSigma}
        where $\sigma_{stat}$ is the statistical error computed as the square root of counts within each radial bin.

        For each sample realization we computed the observed global density profile $n_e^{obs(los)}$ and the observed intrinsic scatter $\sigma_{intr}^{obs(los)}$.
        We then compared different sample realization and considered the mean observed global profile $n_e^{obs}$ of the normal distribution of all the observed global profiles $n_e^{obs(los)}$. 
        The same procedure has been performed on the intrinsic scatter to obtain the mean observed intrinsic scatter $\sigma_{intr}^{obs}$ of all the observed intrinsic scatter $\sigma_{intr}^{obs(los)}$. 
        We also defined the respective input quantities, that is the mean input global density profile $n_e^{in}$ and the mean input intrinsic scatter $\sigma_{intr}^{in}$. 
        
        By comparing observed and input quantities
        we can determine the biases $b_n$ and $b_s$, respectively on density and scatter profiles, associated with the spherical geometry assumption.        
        We define the bias $b_q$ for the quantity $q$ as: 
        \begin{equation}
            b_q = q^{obs}/q^{in} - 1 .
        \end{equation}

 %


    \section{Results} \label{sec:Results}
\subsection{Single Clusters results}\label{sec:4_Results_single}
    \begin{figure*}
        \centering
        \includegraphics[width =0.49\textwidth]{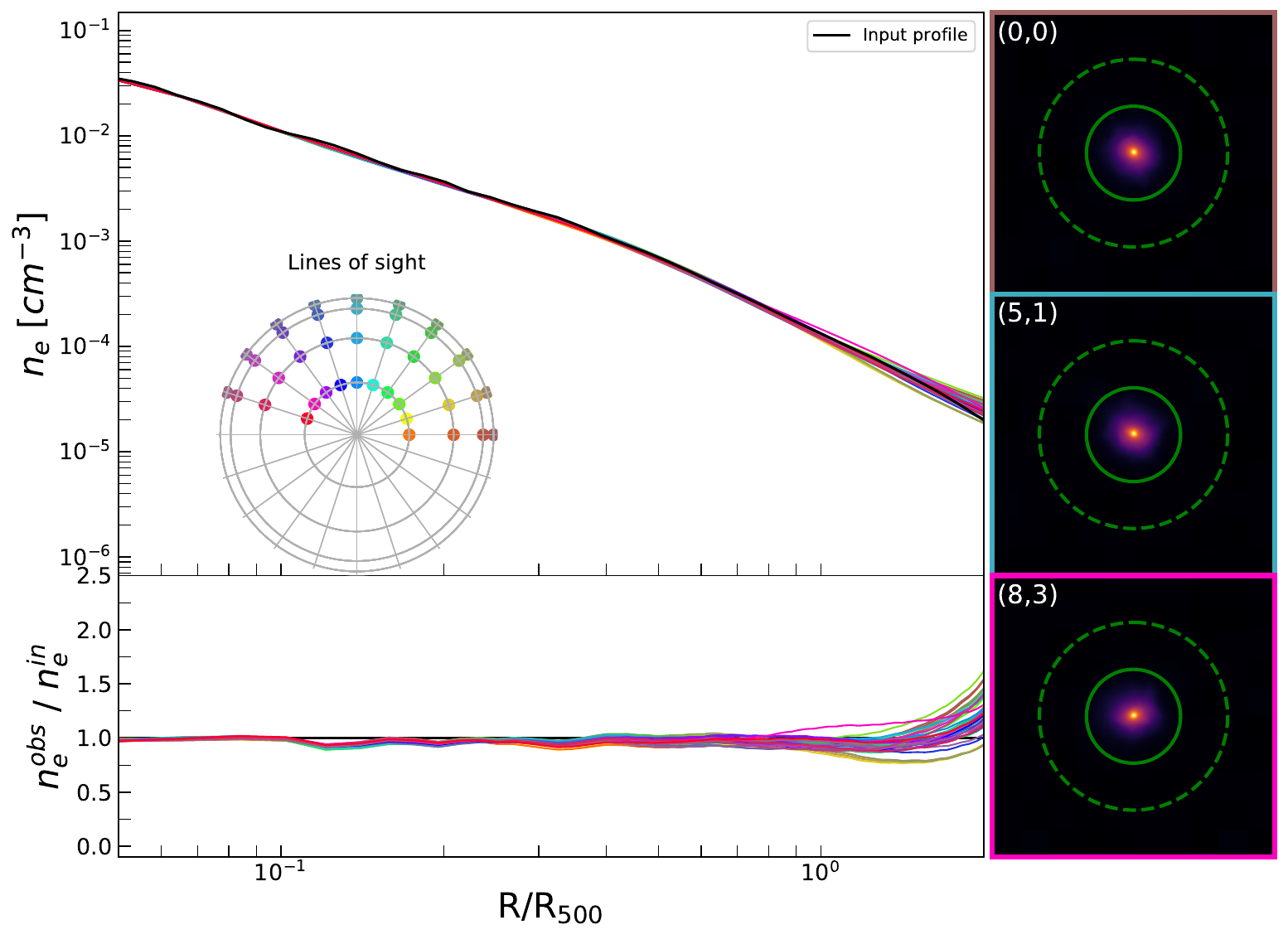}
        \includegraphics[width =0.49\textwidth]{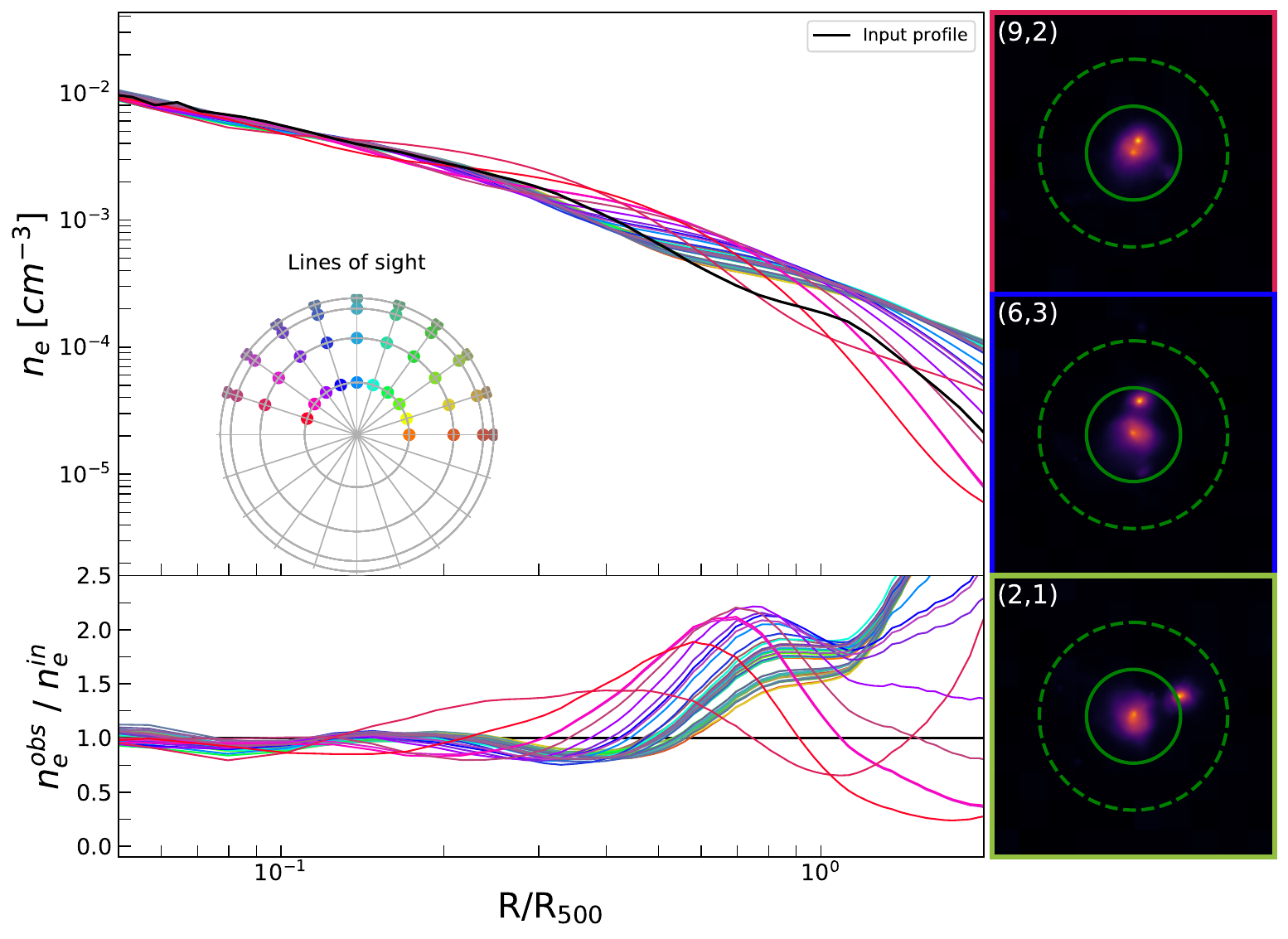}  
        \caption{
           Gas density profile along 40 lines of sight for two clusters chosen as example. In the left figure the regular cluster CL0129\_1 is reported while in the right figure the irregular cluster CL0005\_1.
           For both figures, the black line refers to the input gas density profile and the colored lines refer to the observed profiles, extracted assuming spherical geometry for the gas spatial distribution; each color refers to a specific line of sight, as reported on the polar plane. 
           In the bottom panel of both figures the observed-to-input ratio is reported for each line of sight. It is evident how the input profile is better reconstructed for the regular cluster.
           On the right of each figure, three cluster projections are reported (the edges color identifies the correspondent line of sight). 
           The solid and the dashed circles refer respectively to $R_{500}$ and $2R_{500}$. 
           We can notice the smooth shape of CL0129\_1, regardless of the projection, and the irregular shape of CL0005\_1, where the substructure position changes with the projection.
           } \label{fig:4_single}
    \end{figure*}  
    
    The results of the analysis outlined in Sect.~\ref{sec:3_ProfAnalysis_single} on single clusters show the impact of clusters morphology on the density profile reconstruction. 
    We report two illustrative and opposite cases: one "regular" cluster (i.e. without substructures) and one "irregular" cluster (i.e. with substructures), as shown in the Xray-like maps reported in Fig.~\ref{fig:4_single}.
    
    For the first case we report the cluster \verb|CL0129_1| (Fig.~\ref{fig:4_single}, left): it does not present any substructure, as shown in the three reported maps, and exhibits a spherical core; the cluster appears indeed similar to itself regardless of the considered projection. 
    Therefore, the 40 density profiles should appear similar to each other. 
    Moreover, each reconstructed profile reproduces the input one quite well: the ratio between the extracted and the input profiles is $\lesssim 1.1$ for a very large radial range ($R\lesssim 1.5 R_{500}$). 
    This result indicates the assumed spherical symmetry does not introduce any relevant bias when studying clusters with few substructures.
    
    For the second case we report the irregular cluster \verb|CL0005_1| (Fig.~\ref{fig:4_single}, right). 
    It features an evident substructure visible in all the three cluster projections. Its presence can significantly modify the apparent shape of the cluster that an observer would see. 
    For example, an observer along the line of sight \textit{(9,2)} (the red one in Fig.~\ref{fig:4_single}) might consider this cluster as a regular one (with no substructures in the outer regions), whereas along the \textit{(2,1)} projection (the green one) the substructure is clearly distinguishable from the central core. 
    Because of this, the reconstructed density profiles are expected to differ from one line of sight to another, as actually shown by the results and indeed from the input profile.
    The shape of the reconstructed profile is affected by the substructure position: if we consider once again the projection along \textit{(9,2)}, we can see that the corresponding reconstructed density profile shows a "bump" at small radii, where the substructure is seen: in this region the observed profile results overestimated with respect to the input one.
    If, instead, we consider the \textit{(2,1)} projection (green), we can see that the bump in the density profile shape is located at larger radii, where the substructure appears.
    More generally, for each line of sight, the reconstructed profile overestimates the input profile where the substructure appears. 
    This can be partially due to a projection effect and partially to the spherical modeling process, both of which contribute, making the substructure deprojected density higher than the real three-dimensional one (see Appendix~\ref{app:model}).
    
    From these considerations we infer that for irregular clusters the spherical approximation introduces an overall overestimation of the extracted profile with respect to the input one in the region where the substructures appears, with typical ratios $n_{e, cl}^{obs (los)}/n_{e,cl}^{in} \gtrsim 2 $ for $R \gtrsim R_{500}$. 

    \medskip
    
\subsection{Sample results}\label{sec:4_Results_Global}
\begin{figure*}[t]  
        \centering
        \includegraphics[width =0.49\textwidth]{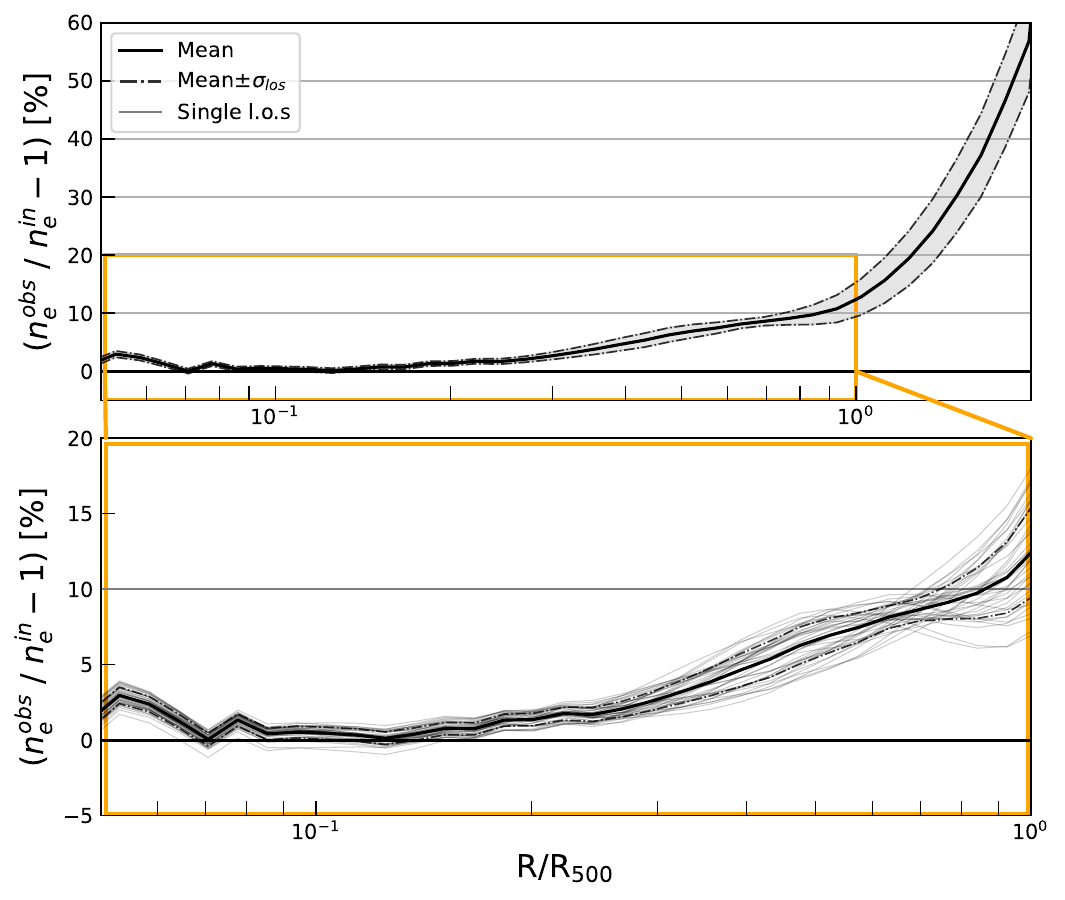}
        \includegraphics[width =0.49\textwidth]{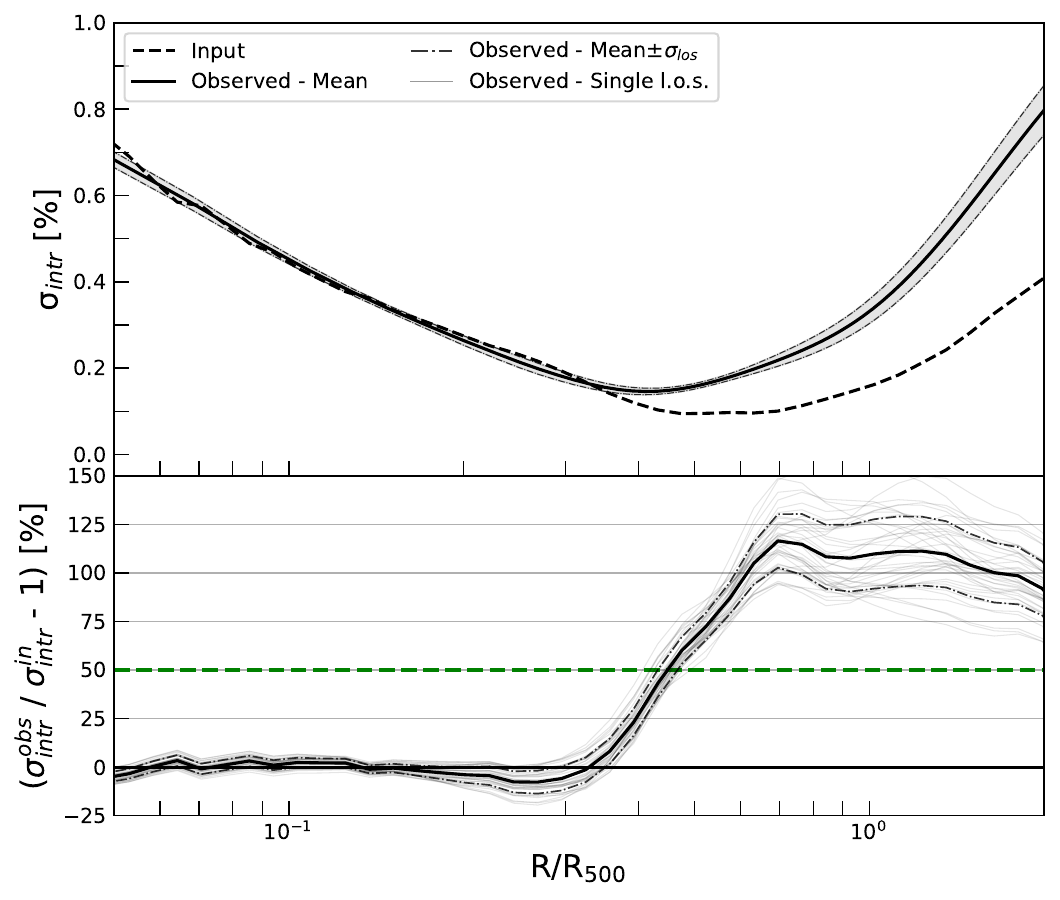}  
        \caption{ Characterisation of the bias introduced by the spherical assumption for the whole sample.
        \textit{Left figure}: bias on ICM density profile due to spherical assumption. In the bottom panel we report the zoom of the innermost region. 
        The grey lines refer to the global density profile for each sample realization (one for each line of sight), the solid black line to the mean over the sample realizations and the dashdotted lines to  $1\sigma$ value. 
        \textit{Right figure}: scatter profile of the sample's density profiles around the global density profile (upper panel) and the relative bias due to spherical assumption (bottom panel). The lighter black lines refer to the scatter profile for each sample realization, the solid black line to the mean over the sample realizations and the dashdotted lines to  $1\sigma$ value; the black dashed line refers to the input profile; the green dashed line to the $50\%$ level. } \label{fig:4_tot}
    \end{figure*}

    The results of the analysis outlined in Sect.~\ref{sec:3_ProfAnalysis_sample} on each sample realization are reported in Fig.~\ref{fig:4_tot}: 
    in the left panel we report the bias $b_n = n_e^{obs}/n_e^{in} - 1$ on the global density profile while in right panel we report the intrinsic scatter $\sigma_{intr}$ profile and its the associated bias $b_s = \sigma_{intr}^{obs}/\sigma_{intr}^{in} - 1$. 
    For both these quantities we report the 40 results on each sample realization ($q^{obs(los)}$) and also the mean over all the projections ($q^{obs}$).
    
    The bias on the global density profile due to the spherical approximation shows a very similar behaviour regardless of the line of sight, exhibiting an overall overestimation of the reconstructed profile that gradually increases with the radius. 
    More specifically, in the innermost regions, for $R \lesssim R_{500}$, the mean bias introduced by the spherical approximation is $\lesssim 10\%$ and decreases down to $\lesssim5\%$ for $R \lesssim 0.4 R_{500}$; conversely, in the outer regions the bias increases, reaching $50\%$ at $R \sim 2R_{500}$ (see Table~\ref{tab:results}).
    As expected, this behaviour does not significantly differ from one sample realization to another, that is, the line of sight does not introduce significant differences in the bias.
    
    These results show that the reconstructed global density profile is generally overestimated. 
    We can hypothesize that this is due to the presence of irregular clusters in the sample, since, as we see in Sect.~\ref{sec:4_Results_single}, the presence of substructures causes an overestimation of the observed cluster emission and thus of the observed density profile in the region where the substructure appears. 
    Therefore, if several irregular clusters are included in the sample, the global density profile should be overestimated over a large radial range, as shown by the results.
    Moreover, the bias increases with the radius. This can be due to the substructures apparent position: the profile tends to be more overestimated if the substructure appears in the outer regions (see Appendix~\ref{app:model}).
            
    We move now to the analysis of the results on the sample intrinsic scatter profile. 
    First of all, we notice that the observed total dispertion $\sigma$ (see Eq.~\ref{eq:3_totSigma})
    almost entirely coincides with the intrinsic scatter $\sigma_{intr}$, as the statistical scatter $\sigma_{stat}$ is very small thanks to the fact that we are using mock observations with high statistics derived from simulations. 
    With these results in hand we observe that the intrinsic scatter (Fig.~\ref{fig:4_tot}, right upper panel) behaviour is similar for both the observed and the input scatter profiles: they show a convex shape with a minimum in $0.4-0.8 R_{500}$; 
    at smaller radii the scatter increases, since simulated clusters present different core densities and slopes;
    in the outer regions the scatter increases, since in these regions clusters more frequently show substructures.
    
    We can evaluate the difference between the observed and the input scatter by analysing the bias $b_s$ (Fig.~\ref{fig:4_tot}, right lower panel).
    We can see that they mostly differ in the outer regions. 
    For $R < 0.3 R_{500}$ the difference between observed and input scatter is in fact not very significant, with an associated bias $b_s < 10\%$. 
    This is due to the fact that in these regions the observed density profile  closely follows its corresponding input profile ($b_{n} \lesssim 5\%$ for $R \lesssim 0.4 R_{500}$) regardless of the cluster and the line of sight, so that the distribution of the observed profiles is very similar to the distribution of the input ones, thus the bias on the scatter is small.
    Conversely, in the outer regions the observed scatter increases much more than the input scatter, with an associated bias $b_s \gtrsim 100 \%$ for $R>0.6 R_{500}$. 
    This is likely due to substructures that show up at different projected radii along different lines of sight, impacting on a large radial range.       

    For both the global density and the intrinsic scatter profiles, we hypothesized that the differences between the observed and the input quantities can be mainly due to the presence of substructures in some clusters. 
    We can test and verify this hypothesis by dividing the sample into sub-samples with different morphology composition.  

    \section{Substructures impact analysis}\label{sec:Discussion}
We tested the substructures impact by defining two sub-samples that we called \textit{regular sample} and \textit{irregular sample}, composed respectively of clusters presenting or free from substructure emission.
Thanks to this differentiation we can isolate the substructures impact and evaluate the related bias.

\subsection{Sub-samples definition}
\begin{figure*}
        \centering
        \includegraphics[width=0.99\textwidth]{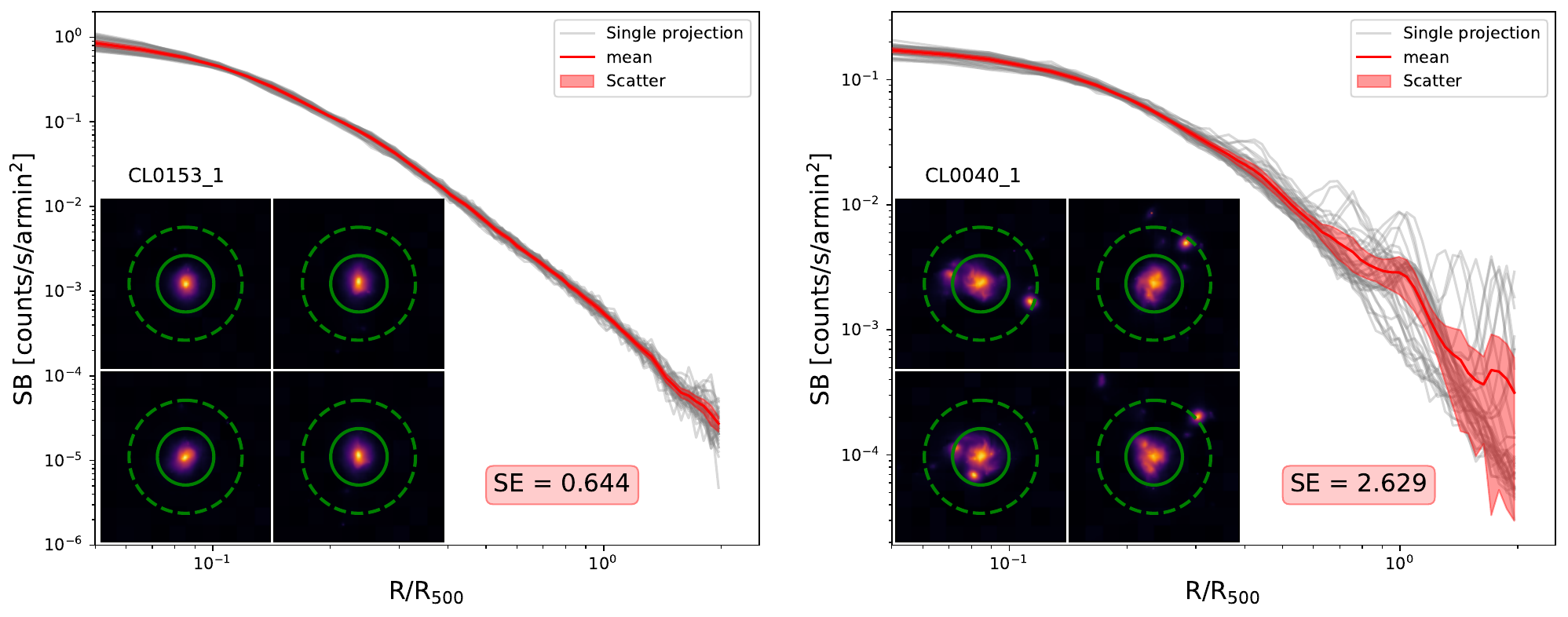}
        \caption{Comparison between each projected surface brightness profiles distribution of a regular (\textit{left} panel) and an irregular (\textit{right} panel) cluster. 
        The regular cluster maps appear very similar to each other despite the projection, so that all the $SB$ profiles referred to different lines of sight present a similar behaviour as well, showing a narrow distribution. 
        On the other hand, the irregular cluster appears very different depending on the considered projection, so that its $SB$ profile presents different shapes, making the profiles distribution scatter to be considerable (depending on the radius). }        \label{fig:5_Reg_Irreg_SB}
    \end{figure*}

    To divide the sample into regular and irregular clusters, we defined a \textit{shape estimator}, that is, a quantity that can evaluate the presence of substructures. 
    Substructures modify the surface brightness profile shape by introducing an $SB$ peak where the substructure appears (for an example see Fig.~\ref{fig:app_sb}). 
    Obviously, the peak position depends on the considered projection. 
    We took advantage of the large number of projection we had and, for each cluster, we compared the surface brightness profiles from different lines of sight: clusters that present substructures will show different surface brightness profile shapes depending on the considered projection, while regular clusters will show very similar profiles regardless of the line of sight (Fig.~\ref{fig:5_Reg_Irreg_SB}). 
    Therefore, we can distinguish between regular and irregular clusters by evaluating the surface brightness profiles distribution width, normalized for the mean surface brightness value, i.e. their scatter $\sigma_{SB}$. 

   There are several morphological indicators that are used in the literature to classify clusters that have been calibrated on X-ray observations (e.g. see \citealt{campitiello2022} for a recent work using these indicators). However, their application on our sample of simulated X-ray observations would require further calibrations which are beyond the scope of this work. For this reason, we defined the shape estimator $SE$ as the combination of different profiles distribution scatters at different radii: 
    \begin{equation}
        SE = \sqrt{\sum_i{\sigma^2_{SB}(r_i)}}.
    \end{equation}
    
    In particular, we considered seven radii between $0.1 R_{500}$ and $2.0R_{500}$ (see Appendix~\ref{app:SE}).
    We then computed the $SE$ for all the clusters in the sample and ordered them in ascending order with respect to the $SE$ value, so that the first clusters are the most regular ones, while the last are the most irregular ones. 
    We have identified two sub-samples, one of regular and one of irregular clusters, by selecting respectively the first and the last 15 $SE$-ordered clusters (see Fig.~\ref{fig:app_se_ord}).   

    The regular clusters are characterized by the lack of substructures, however their morphologies typically differ from a spherical shape, showing triaxial structures. 
    Thanks to this characteristic, we can use the regular sample to study the spherical assumption bias on non-spherical objects without substructures, i.e. without the main source of deviation from the spherical symmetry. 

\subsection{Sub-samples analysis}
\begin{figure*}[t]  
        \centering
        \includegraphics[width=0.49\textwidth]{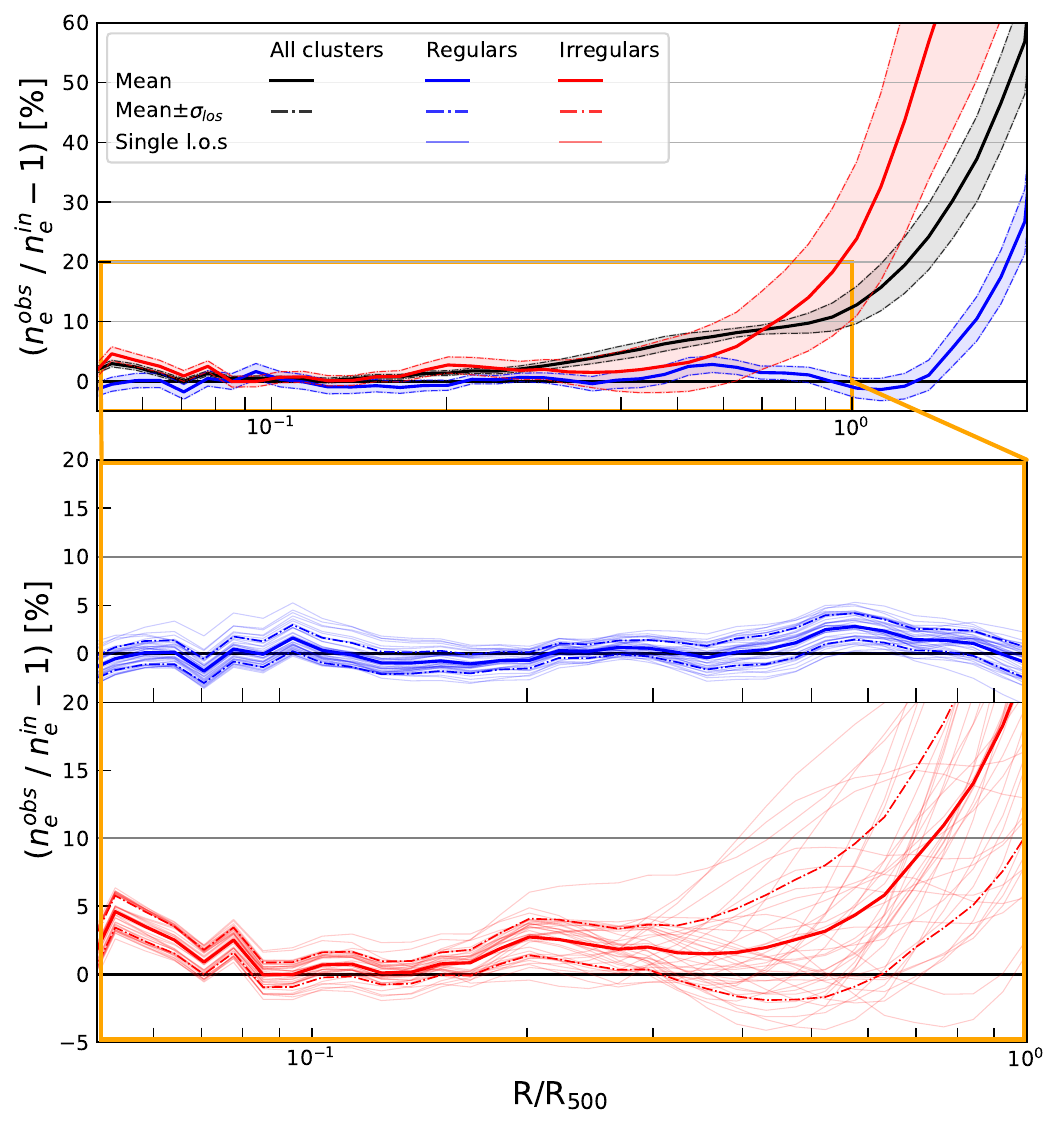}
        \includegraphics[width=0.49\textwidth]{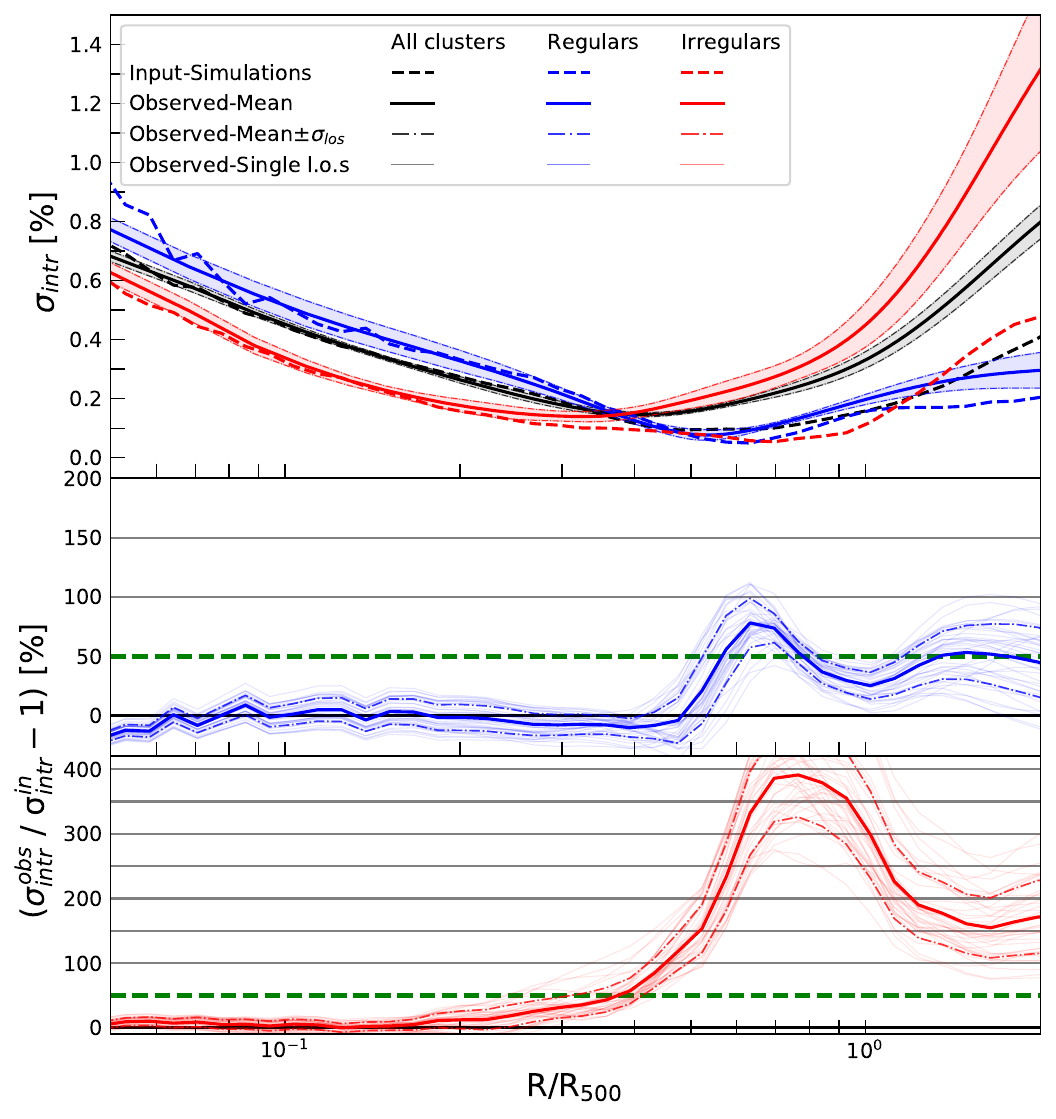}    
        \caption{
        Same as Figure~\ref{fig:4_tot} for the regular and irregular sub-samples. 
        \textit{Left figure}: bias on ICM density profile due to spherical assumption. In the bottom panel the zoom of the innermost region is reported. 
        \textit{Right figure}: scatter profile of each sample's density profiles around the global density profile (upper panel) and the relative bias due to spherical assumption (bottom panel).
        The black lines refer to the complete sample, the red lines to to irregular sample and the blue lines to the regular sample.     
        The lighter lines refer to the global density profile for each sample realization (one for each line of sight), the solid lines to the mean over the sample realizations and the dashdotted lines to  $1\sigma$ value. 
        In the right figure the dashed lines refer to the input profile; the green dashed line to the $50\%$ level.}
        \label{fig:5_SubSam}
    \end{figure*}

    We repeated the same analysis outlined in Sect.~\ref{sec:3_ProfAnalysis_sample} on the two sub-samples just defined, obtaining for each sub-sample 40 global density profiles $n_{e,I/R}^{obs(los)}$ and 40 global scatter profiles $\sigma_{intr,I/R}^{obs(los)}$ (one for each sample realization along a line of sight) and the mean global density and scatter profiles over the 40 sample realizations, $n_{e,I/R}^{obs}$ and $\sigma_{intr,I/R}^{obs}$.
    The results are reported in Fig.~\ref{fig:5_SubSam} and Table~\ref{tab:results}, compared with the results obtained in Sect.~\ref{sec:4_Results_Global} for the complete sample. 
    As we can see, the sample composition strongly affects both the reconstructed density and scatter profiles. 

    For what concerns the density profile, we can see that for the regular clusters sample the observed profile differs from the input profile less than $10\%$ up to very large radial range ($R\approx 1.5 R_{500}$) while for the complete sample the $10\%$-bias range is achieved only up to $R \approx 0.9 R_{500}$.
    In addition, for the regular sample the bias remains $\lesssim 5\%$ for a very extended radial range ($R\lesssim 1.2 R_{500}$).
    On the other hand, for the irregular clusters sample, the observed density profile is well reconstructed ($b_n< 10 \%$) for a smaller radial range ($R < 0.7 R_{500}$) while at $R \approx R_{500}$ the observed profile is overestimated by more than $20\%$ with respect to the input profile.
    
    We can conclude that the differences between the observed and the input profiles strongly depend on the sample composition and in particular on its large-scale morphology. That is, if the sample is composed of clusters that present substructures, then the density profile of the sample is less accurately reconstructed than for regular clusters samples.
    This is highlighted also by the radius at which we see the difference between the reconstructed and the input profile: for irregular clusters the bias increases significantly at $R \approx R_{500}$, where indeed substructures begin to appear.
    Considering the complete sample, where both regular and irregular clusters are present, the substructures effects are reduced by the regular clusters, so that the global profile of the sample is better reconstructed for a large radial range than the case where only irregular clusters are considered. 

    Differences between the regular and irregular samples become even more pronounced when we investigate the scatter.
    The regular sample presents an observed scatter much more similar to the input one than the complete sample, with $b_s \lesssim 10\%$ for $R \lesssim 0.5 R_{500}$. The bias reaches a maximum value of $\approx 90\%$ for $0.6 R_{500} \lesssim R \lesssim 0.7 R_{500}$ and then stabilizes to $\approx 50\%$ for $R > R_{500}$. 
    For the irregular sample we can see that the observed scatter differs from the input one much more than the complete and the regular samples, even for small radii, being $\lesssim 10\%$ only for $R \lesssim 0.2 R_{500}$, while at $R \approx 0.3 R_{500}$ we find $b_s \approx 40\%$. The bias increases up to very high values in the outer regions: we find a maximum bias of $\approx 400\%$ for $0.7 R_{500} \lesssim R \lesssim R_{500}$ and $b_s \approx 200\%$ for $R > R_{500}$.
    
    These results confirm that the observed density profiles of irregular clusters result overestimated in the radial range where the substructures appear: since the substructures position differs from one cluster to another, the sample profile distribution will be broader than the input one. 
    For regular cluster this is obviously not valid: since there are no substructures, the density profile is well reconstructed for all the clusters in the sample, making the sample profile distribution much more similar to the input one. 

    The substructures impact is also noticeable if we study the results for different sample realizations. 
    For the regular sample every line of sight shows a very similar behaviour, while in the irregular sample important differences can be found from one sample realization to another. This is obviously due to substructures: in the regular sample each cluster appears similar to itself regardless of the line of sight, so that there are not large differences between the sample realizations; on the other hand, in the irregular sample each cluster can appear in very different shapes depending on the considered line of sight, creating very different sample realizations (see Fig.~\ref{fig:app_subsample}).
    
    These results lead to the conclusion that the high bias measured for the complete sample is indeed due to substructures. In fact, since the sample is composed of both regular and irregular clusters, the profile distribution is primarily enlarged by irregular clusters. 
    
    \renewcommand{\arraystretch}{1.2}
    \begin{table*}[t]
        \centering
        \caption{Results of the analysis for the biases on the global density profile and on the intrinsic scatter of a cluster population. } \label{tab:results}
        \begin{tabular}{c|ccc|ccc}
           \toprule \toprule
           
                & \multicolumn{3}{c|}{Bias on global density} 
                    & \multicolumn{3}{c}{Bias on density scatter [$\%$]}          \\
                    
                & \multicolumn{3}{c|}{$b_n \pm \sigma_n^{los}$ [$\%$]} 
                    & \multicolumn{3}{c}{$b_s \pm \sigma_s^{los}$ [$\%$]}          \\
            R [$R_{500}$]
                & Complete sample & Regular sample & Irregular sample 
                    & Complete sample & Regular sample & Irregular sample  \\
           \hline
            0.2 	& 1.37 $\pm$ 0.01 	& $-0.66$ $\pm$ 0.02 	& 2.74 $\pm$ 0.03 	& -3.9 $\pm$ 0.1 	& -1.3 $\pm$ 0.3 	& 12.4 $\pm$ 0.4 \\
            
            0.5 	& 6.61 $\pm$ 0.03 	& 1.82 $\pm$ 0.04 	& 2.9 $\pm$ 0.1 	& 67.1 $\pm$ 0.2 	& 9.3 $\pm$ 0.6 	& 139.6 $\pm$ 0.8 \\
            
            0.8 	& 9.43 $\pm$ 0.03 	& 1.24 $\pm$ 0.03 	& 12.5 $\pm$ 0.2 	& 111.1$\pm$ 0.4 	& 46.9 $\pm$ 0.3 	& 374.1 $\pm$ 1.6 \\
            
            1.0 	& 12.8 $\pm$ 0.08 	& -1.14 $\pm$ 0.04 	& 23.9 $\pm$ 0.3 	& 110.1 $\pm$ 0.5 	& 25.8 $\pm$ 0.3 	& 287.7 $\pm$ 1.6 \\
            
            1.5 	& 30.3 $\pm$ 0.2 	& 5.63 $\pm$ 0.08 	& 67.0 $\pm$ 0.7 	& 104.9 $\pm$ 0.4 	& 55.2 $\pm$ 0.6 	& 159.9 $\pm$ 1.1 \\
            
            2.0 	& 56.9 $\pm$ 0.2 	& 26.7 $\pm$ 0.1 	& 105.3 $\pm$ 0.9 	& 95.4 $\pm$ 0.4 	& 45.2 $\pm$ 0.7 	& 174.1 $\pm$ 1.4 \\
           \bottomrule
        \end{tabular}
        \vspace{1ex}
        
        {\raggedright \textbf{Notes}. The results are given for the entire sample (98 clusters) and for two sub-samples (15 clusters each one) composed of regular or irregular clusters. The errors on mean values are attributed to the differences given by different sample realizations along different lines of sight: for the quantities $q$, $\sigma_q^{los} = ({b_{q}^{los}}_{MAX} + {b_{q}^{los}}_{MIN})/2$.\par }

    \end{table*}

    
    \section{Conclusions} \label{sec:Conclusion}
In this paper we tested the three-dimensional reconstruction of the global ICM density profile of a cluster population and its intrinsic scatter profile when spherical geometry is assumed.
We used a 98 simulated galaxy clusters sample that was built to be representative of a common cluster population.
We considered 40 different sample realizations by projecting each cluster along 40 different lines of sight. 
For each sample realization we emulated an X-ray observation of all the clusters in the sample and we extracted the electron density profile assuming a spherical geometry for the gas spatial distribution. 
We then derived the global gas density profile for each sample realizations and compared it with the input global density profile, directly given by the simulations. 
Therefore, we were able to determinate the bias introduced by the spherical assumption on the reconstruction of the ICM density profile.
We also analyse the bias on the intrinsic scatter of the profile distribution for each sample realization. 

\medskip 

\noindent For the global density profile we found that: 
\begin{itemize}
    \item it is well reconstructed for a large radial range, with a bias due to the spherical approximation smaller than $10\%$ for $R \lesssim R_{500}$;
    \item in the innermost regions ($R\lesssim 0.4 R_{500}$) the bias decreases down to values $\lesssim 5\%$;
    \item at large radii ($R>R_{500}$) the bias increases up to $\approx 50\%$ because of the impact of substructures;
    \item the bias strongly depends on the sample composition: if the sample is composed of regular clusters (without overdensity substructures) the bias is smaller ($b_n\lesssim 20\%$ for $R\lesssim 2 R_{500}$ and $b_n< 5\%$ for $R\lesssim 1.5 R_{500}$), while if it is only composed of clusters that show substructures the bias considerably increases ($b_n>10\%$ for $R \gtrsim 0.7 R_{500}$). 
\end{itemize}

\medskip

\noindent For the intrinsic scatter of the density profiles distribution we found that: 
\begin{itemize}
    \item the mean reconstructed scatter follows the input one quite well in the inner regions, with bias $b_s < 10\%$ for $R \lesssim 0.3 R_{500}$; 
    \item in the outer regions the reconstructed scatter results considerably higher than the input one, with an associated bias that rapidly increases up to $\approx 100\%$ in $0.4 R_{500}\lesssim R \lesssim 0.7 R_{500}$, maintaining this value for larger radii. This is due to substructures in the outer regions of some clusters in the sample, that cause the density profile to be overestimated in the radial range where substructures appear, making the sample profile distribution broader; 
    \item the bias on scatter strongly depends on the sample composition, as for the bias on density profile: 
    if we consider a sample composed exclusively of clusters that do not show substructures, the bias on the profile distribution scatter is $b_s < 10\%$ for a larger radial range ($R \lesssim 0.5 R_{500}$) and in the outer regions it is reduced by a factor $2$, down to $\approx 50\%$; 
    if we consider a sample composed exclusively of clusters that do show substructures, the bias is $\lesssim 10\%$ only for $R \lesssim 0.2 R_{500}$ (at $R \approx 0.3 R_{500}$ we find $b_s \approx 40\%$); in the outer regions it increases up to a maximum value of $\approx 400\%$ for $0.7 R_{500} \lesssim R \lesssim R_{500}$ and $b_s \approx 200\%$ for $R > R_{500}$. 
\end{itemize}

The analysis outlined in this paper led to the estimation of the biases that should be considered when 3D reconstruction of ICM density profile through spherical approximation is performed on real observations, in particular considering cluster substructures. 
In fact, even if they are generally masked out from real observations, it can append that some of them are not distinguishable and consequently they can not be masked, introducing the bias that we evaluated in this work. 
Moreover, the shape of the intrinsic scatter found in the analysis can be used as a comparison for real observations on clusters samples.

    
    \begin{acknowledgements} 
        The authors thank the anonymous referee for the useful comments that improved the
quality of the manuscript.
        We acknowledge financial contribution from the contracts ASI-INAF Athena 2019-27-HH.0,
        ``Attivit\`a di Studio per la comunit\`a scientifica di Astrofisica delle Alte Energie e Fisica Astroparticellare''
        (Accordo Attuativo ASI-INAF n. 2017-14-H.0), and
        from the European Union’s Horizon 2020 Programme under the AHEAD2020 project (grant agreement n. 871158).
        This work has been made possible by the ‘The Three Hundred’ collaboration, which received financial support from the European Union’s Horizon 2020 Research and Innovation programme under the Marie Sklodowskaw-Curie grant agreement number 734374, i.e. the LACEGAL project. The simulations used in this paper have been performed in the MareNostrum Supercomputer at the Barcelona Supercomputing centre, thanks to CPU time granted by the Red Española de Supercomputación.
     \end{acknowledgements}

    \bibliographystyle{aa}
    \bibliography{biblio}

    \begin{appendix}
        \section{Sample properties} \label{app:tabel}
\begin{minipage}{0.93\textwidth}
\renewcommand{\arraystretch}{1.1}
    \captionof{table}{Table of the sample properties.} \label{tab:A_1}
    \begin{tabular}{lllll}
        \toprule \toprule
        \multirow{2}{*}{Cluster}  & \multicolumn{1}{c}{$R_{500}$} & \multicolumn{1}{c}{$M_{500}$}  
                        & \multicolumn{1}{c}{$T_{500}$}   	        &  \multirow{2}{*}{$SE_{tot}$} \\
                        
        			           &  \multicolumn{1}{c}{[arcmin]} & \multicolumn{1}{c}{[$10^{14} \mathrm{M}_\odot$]} 
                        &  \multicolumn{1}{c}{[$\mathrm{keV}$]} & 		 \\
        \midrule 
        CL0001\_1 	& 4.271 	& 7.670 	& 5.755 	& 2.101  	 \\
        CL0001\_2 	& 3.429 	& 3.969 	& 4.152 	& 0.867  	 \\
        CL0001\_3 	& 3.594 	& 4.571 	& 4.467 	& 1.632  	 \\
        CL0003\_1 	& 4.431 	& 8.567 	& 6.515 	& 0.837  	 \\
        CL0003\_3 	& 3.828 	& 5.522 	& 3.888 	& 0.688  	 \\
        CL0004\_2 	& 3.831 	& 5.536 	& 5.018 	& 0.498 (R)	 \\
        CL0005\_1 	& 3.661 	& 4.831 	& 4.175 	& 1.872  	 \\
        CL0006\_2 	& 3.342 	& 3.674 	& 4.154 	& 0.670  	 \\
        CL0006\_3 	& 2.774 	& 2.101 	& 2.912 	& 1.323  	 \\
        CL0008\_2 	& 3.052 	& 2.798 	& 2.926 	& 0.945  	 \\
        CL0008\_3 	& 2.745 	& 2.036 	& 2.090 	& 0.609 (R)	 \\
        CL0009\_1 	& 5.256 	& 14.291 	& 10.107 	& 1.268  	 \\
        CL0009\_2 	& 3.033 	& 2.748 	& 3.739 	& 0.601 (R)	 \\
        CL0010\_1 	& 3.570 	& 4.480 	& 3.780 	& 1.593  	 \\
        CL0010\_2 	& 3.517 	& 4.282 	& 4.679 	& 0.774  	 \\
        CL0010\_3 	& 3.328 	& 3.630 	& 3.855 	& 0.537 (R)	 \\
        CL0012\_1 	& 3.915 	& 5.910 	& 3.849 	& 6.772 (I)  \\
        CL0013\_1 	& 4.145 	& 7.009 	& 5.013 	& 1.030  	 \\
        CL0014\_1 	& 2.923 	& 2.458 	& 2.881 	& 2.005  	 \\
        CL0015\_1 	& 3.800 	& 5.402 	& 4.662 	& 3.411 (I)  \\
        CL0016\_3 	& 3.256 	& 3.398 	& 3.794 	& 2.287 (I)	 \\
        CL0017\_1 	& 4.699 	& 10.215 	& 7.337 	& 0.698  	 \\
        CL0018\_1 	& 4.860 	& 11.297 	& 8.446 	& 1.057  	 \\
        CL0018\_3 	& 3.264 	& 3.423 	& 4.086 	& 1.741  	 \\
        CL0020\_3 	& 3.782 	& 5.324 	& 4.660 	& 0.769  	 \\
        CL0021\_1 	& 4.685 	& 10.124 	& 6.391 	& 1.702  	 \\
        CL0022\_2 	& 3.219 	& 3.285 	& 4.700 	& 0.810  	 \\
        CL0025\_2 	& 3.044 	& 2.764 	& 3.040 	& 0.833  	 \\
        CL0025\_1 	& 4.438 	& 8.604 	& 8.218 	& 0.920  	 \\
        CL0026\_1 	& 4.288 	& 7.763 	& 5.519 	& 2.512 (I)	 \\
        CL0026\_2 	& 3.444 	& 4.022 	& 4.469 	& 0.546 (R)	 \\
        CL0026\_3 	& 3.031 	& 2.740 	& 3.492 	& 1.200  	 \\
        CL0026\_4 	& 2.739 	& 2.022 	& 2.630 	& 0.496 (R)  \\
        CL0027\_1 	& 4.775 	& 10.716 	& 7.873 	& 0.664  	 \\
        CL0028\_1 	& 3.801 	& 5.408 	& 5.641 	& 1.984  	 \\
        CL0029\_1 	& 4.998 	& 12.293 	& 8.152 	& 1.297  	 \\
        CL0034\_1 	& 3.820 	& 5.489 	& 4.162 	& 1.065  	 \\
        CL0034\_2 	& 2.880 	& 2.351 	& 3.167 	& 0.544 (R)  \\
        CL0035\_1 	& 4.633 	& 9.792 	& 5.781 	& 2.413 (I)  \\
        CL0036\_1 	& 4.376 	& 11.514 	& 7.428 	& 0.788  	 \\
        CL0039\_1 	& 4.821 	& 11.031 	& 7.022 	& 1.715  	 \\
        CL0040\_1 	& 4.233 	& 7.466 	& 5.255 	& 2.629 (I)  \\
        CL0042\_2 	& 3.055 	& 2.805 	& 2.626 	& 1.303  	 \\
        CL0043\_1 	& 4.349 	& 8.095 	& 5.401 	& 1.085  	 \\
        CL0046\_1 	& 4.504 	& 8.995 	& 6.966 	& 10.129 (I) \\
        CL0047\_1 	& 3.714 	& 5.042 	& 3.683 	& 1.284  	 \\
        CL0047\_2 	& 3.494 	& 4.198 	& 3.728 	& 0.846  	 \\
        CL0047\_3 	& 3.331 	& 3.637 	& 3.827 	& 0.843  	 \\
        CL0049\_1 	& 3.399 	& 3.865 	& 3.430 	& 1.136  	 \\
        \bottomrule
        \bottomrule
    \end{tabular}
    \hfill
    \begin{tabular}{lllll}
        \toprule \toprule
         \multirow{2}{*}{Cluster}  & \multicolumn{1}{c}{$R_{500}$} & \multicolumn{1}{c}{$M_{500}$}  & \multicolumn{1}{c}{$T_{500}$}   	        &  \multirow{2}{*}{$SE_{tot}$} \\
        			            &  \multicolumn{1}{c}{[arcmin]} & \multicolumn{1}{c}{[$10^{14} \mathrm{M}_\odot$]} &  \multicolumn{1}{c}{[$\mathrm{keV}$]} & 		 \\
        \midrule
        CL0049\_2 	& 3.101 	& 2.935 	& 3.507 	& 0.851  	 \\
        CL0053\_1 	& 4.330 	& 7.993 	& 5.387 	& 0.896  	 \\
        CL0055\_1 	& 4.403 	& 8.405 	& 7.408 	& 1.945  	 \\
        CL0057\_1 	& 3.330 	& 3.634 	& 4.412 	& 3.573 (I)	 \\
        CL0059\_1 	& 4.487 	& 8.890 	& 5.695 	& 2.153 (I)	 \\
        CL0061\_1 	& 3.416 	& 3.925 	& 3.784 	& 0.584 (R)	 \\
        CL0061\_2 	& 3.296 	& 3.524 	& 3.143 	& 0.889  	 \\
        CL0067\_1 	& 4.806 	& 10.928 	& 7.186 	& 1.154  	 \\
        CL0068\_1 	& 4.355 	& 8.127 	& 4.713 	& 1.828  	 \\
        CL0069\_1 	& 4.891 	& 8.247 	& 6.621 	& 3.311 (I)	 \\
        CL0072\_1 	& 4.321 	& 7.946 	& 6.479 	& 1.346  	 \\
        CL0074\_1 	& 3.302 	& 3.546 	& 2.998 	& 2.479 (I)	 \\
        CL0075\_1 	& 4.390 	& 8.328 	& 6.990 	& 0.927  	 \\
        CL0076\_1 	& 3.723 	& 5.080 	& 4.076 	& 2.307 (I)	 \\
        CL0079\_1 	& 4.487 	& 8.889 	& 5.746 	& 1.230  	 \\
        CL0081\_1 	& 3.801 	& 5.405 	& 4.626 	& 0.664  	 \\
        CL0087\_1 	& 3.570 	& 4.478 	& 4.627 	& 0.467 (R)	 \\
        CL0089\_1 	& 3.544 	& 4.380 	& 4.376 	& 2.243 (I)	 \\
        CL0091\_1 	& 3.718 	& 5.061 	& 4.307 	& 1.577  	 \\
        CL0092\_1 	& 4.183 	& 7.205 	& 4.403 	& 1.590  	 \\
        CL0093\_1 	& 3.421 	& 3.941 	& 3.389 	& 3.261 (I)	 \\
        CL0094\_1 	& 3.371 	& 3.770 	& 3.617 	& 2.036  	 \\
        CL0098\_2 	& 3.494 	& 4.198 	& 3.861 	& 0.633  	 \\
        CL0099\_1 	& 4.362 	& 8.169 	& 5.523 	& 1.231  	 \\
        CL0102\_1 	& 4.486 	& 8.886 	& 7.751 	& 0.627 (R)	 \\
        CL0105\_1 	& 3.535 	& 4.350 	& 2.917 	& 1.506  	 \\
        CL0107\_1 	& 4.259 	& 7.604 	& 5.401 	& 1.355  	 \\
        CL0110\_1 	& 4.350 	& 8.105 	& 4.994 	& 1.551  	 \\
        CL0119\_1 	& 4.242 	& 7.515 	& 4.554 	& 3.060 (I)	 \\
        CL0120\_1 	& 4.414 	& 8.467 	& 5.898 	& 0.911  	 \\
        CL0129\_1 	& 4.474 	& 8.817 	& 8.549 	& 0.381 (R)	 \\
        CL0134\_1 	& 4.387 	& 8.310 	& 6.085 	& 0.746  	 \\
        CL0146\_1 	& 4.367 	& 8.198 	& 4.381 	& 1.771  	 \\
        CL0153\_1 	& 4.422 	& 8.510 	& 6.741 	& 0.644  	 \\
        CL0154\_1 	& 4.198 	& 7.285 	& 3.555 	& 1.126  	 \\
        CL0155\_1 	& 4.237 	& 7.487 	& 4.708 	& 0.998  	 \\
        CL0160\_1 	& 4.218 	& 7.385 	& 6.918 	& 0.756  	 \\
        CL0189\_1 	& 4.263 	& 7.626 	& 5.327 	& 0.564 (R)	 \\
        CL0191\_1 	& 4.251 	& 7.561 	& 4.120 	& 1.019  	 \\
        CL0206\_1 	& 4.273 	& 7.680 	& 7.187 	& 0.552 (R)  \\
        CL0213\_1 	& 4.231 	& 7.458 	& 4.557 	& 1.009  	 \\
        CL0217\_1 	& 4.356 	& 8.138 	& 6.197 	& 0.992  	 \\
        CL0226\_1 	& 4.180 	& 7.191 	& 6.133 	& 0.743  	 \\
        CL0249\_1 	& 4.302 	& 7.838 	& 5.043 	& 1.497  	 \\
        CL0254\_1 	& 4.285 	& 7.745 	& 5.691 	& 0.912  	 \\
        CL0268\_1 	& 4.319 	& 7.931 	& 4.982 	& 1.525  	 \\
        CL0272\_1 	& 4.277 	& 7.701 	& 6.567 	& 0.592 (R)	 \\
        CL0276\_1 	& 4.168 	& 7.127 	& 7.426 	& 0.455 (R)  \\
        CL0286\_1 	& 4.313 	& 7.899 	& 5.115 	& 0.870  	 \\
        \bottomrule
        \bottomrule
    \end{tabular}
     \vspace{1ex}
     
    {\raggedright \textbf{Note}. The forth column reports the shape estimator value (Appendix~\ref{app:SE}) and the cluster classification: (R) are clusters in the regular sub-sample, (I) are clusters in the irregular sub-sample.\par
    }
\end{minipage}
\clearpage
        \newpage
        \section{Substructures impact}\label{app:model}
\begin{figure*}[b]  
        \centering
        \includegraphics[trim={0 4.5cm 0 4cm},clip, width =\textwidth]{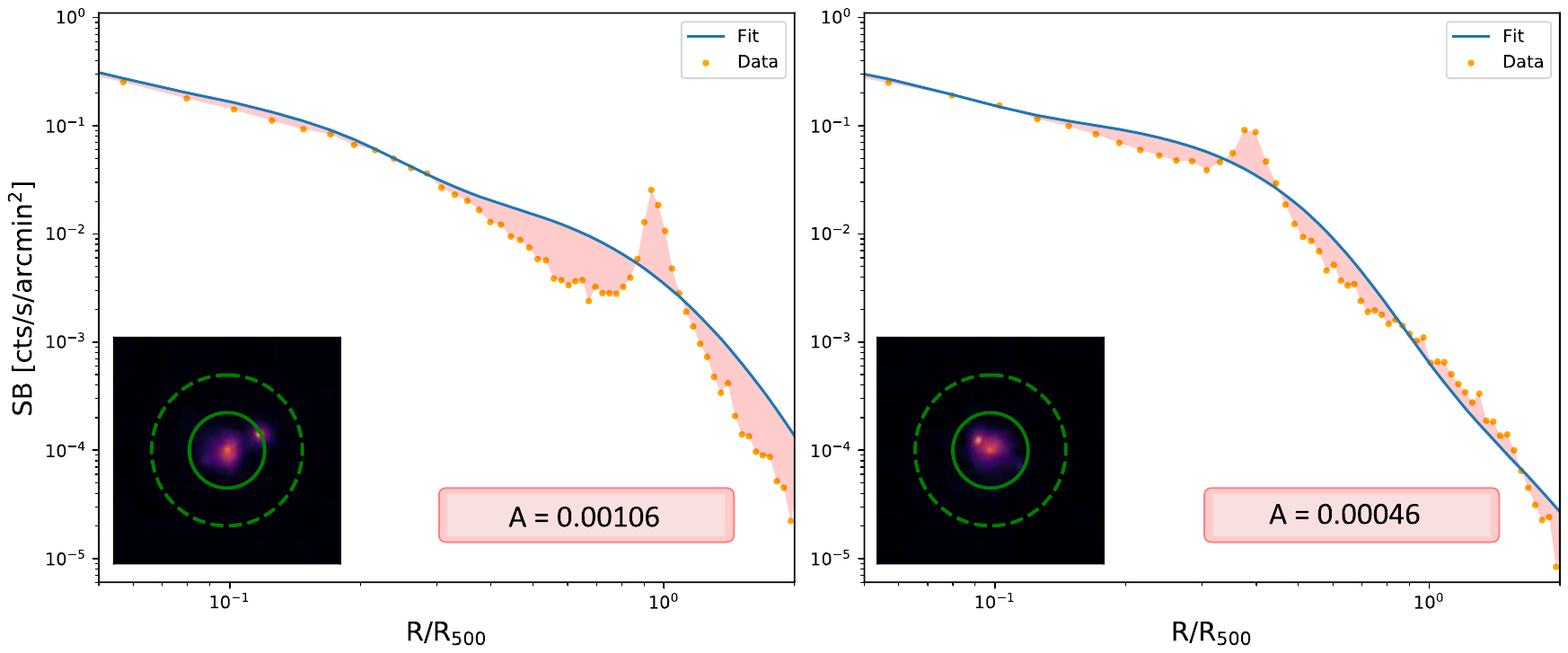}
        \caption{Surface brightness profile for cluster CL0005\_1 seen from two different line of sight. The orange data point refers to the measured emission, the blue line refers to the fitted $\beta$-profiles combination model. 
        The emission peaks correspond to the substructure emission: in the projection reported in the left figure the substructure appears in the outer regions of the cluster ($R \sim R_{500}$), while in the projection reported on the right the substructure appears near the central core. 
        The difference between the fitted profile and the observed one can be quantified by the colored area that is larger when the substructure appears far from the central core,
        \NEW{which means that if the substructure appear far from the central core the observed density profile results much more overestimated.}}
        \label{fig:app_sb}
\end{figure*}  

In our study the observed density profiles result generally overestimated with respect to the input profile.
In particular, the overestimation arise in the region where the substructure appears.
This result can be justified considering two effects: the projection effect and the model effect. 

The density profile derives in fact from the measured surface brightness profile. 
The observed surface brightness profile is obtained by measuring the projected gas emission averaged on annular bins, so that the three-dimensional emission is redistributed on circular shells instead of spherical shells. 
Since surface brightness is related to square density as $SB \propto \langle n^2 \rangle$, if inhomogeneities are present in the gas distribution their emission is enhanced.
This can cause an overestimation of the gas density profile in the order of $\sqrt{C}$ where $C = \langle n^2 \rangle / \langle n \rangle^2$ is the clumpiness factor \citep[see e.g.][]{Nagai, Roncarelli, Ecket_clump}. 
These effects make substructures appear as high emission peak. 

Moreover, the model we used does not take into account the substructures emission peak. 
In fact, we are assuming a spherical shape of the gas distribution, so that the observed emission measure is modelled as a composition of $\beta$-profiles (Eq.~\ref{eq:3_EM_beta}) that can't describe the emission peak. 
The resulting fitted surface brightness profile is forced to pass through the peak, resulting less steep than the measured one in the region around the substructure, leading to an overestimation of the real profile (Figure~\ref{fig:app_sb}). 
As a consequence, the reconstructed density profile results overestimated as well.

Moreover, as shown in Section~\ref{sec:4_Results_Global} and Section~\ref{sec:4_Results_single} the overestimation of the density profiles increases gradually with the radius, in particular in the complete sample and for the irregular clusters, showing that this effect is due to substructures.
This radial variation of the density bias can be explained through the two effects just outlined. 
In fact, if we consider a projection where the substructure appears far from the central core, where the cluster main emission is lower (Figure~\ref{fig:app_sb}, left), the substructure emission peak emerges from the main emission profile much more than the case when the same substructure appears near the central core (Figure~\ref{fig:app_sb}, right). 
In this way, the overestimation of the surface brightness, thus the overestimation of the density profile, is higher for larger radii, as shown by the colored areas in Figure~\ref{fig:app_sb}.  
Therefore, if we consider a sample in which substructures can appear at any distance from the core, the bias on the global density profile does increase with the radius.


        \section{Shape estimator} \label{app:SE}
\begin{figure*}[t]  
        \centering
        \includegraphics[width =\textwidth]{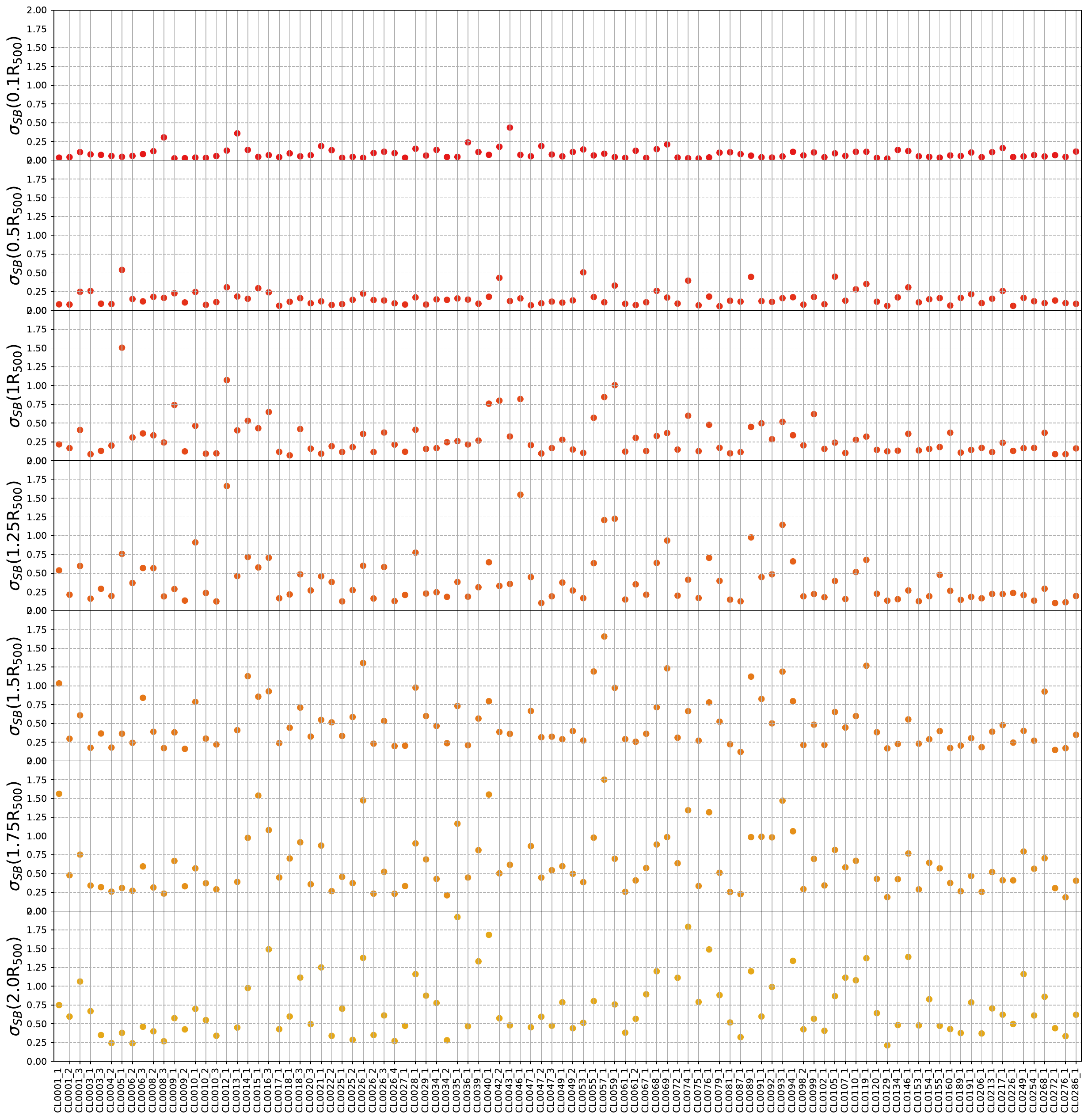}
        \caption{Scatter of the surface brightness profiles distribution (over the 40 line of sight) for each cluster at different radii. The scatters distribution becomes broader with increasing radius.
        }
        \label{fig:app_se_i}
\end{figure*}  

\begin{figure*}[t]  
        \centering
        \includegraphics[width =\textwidth]{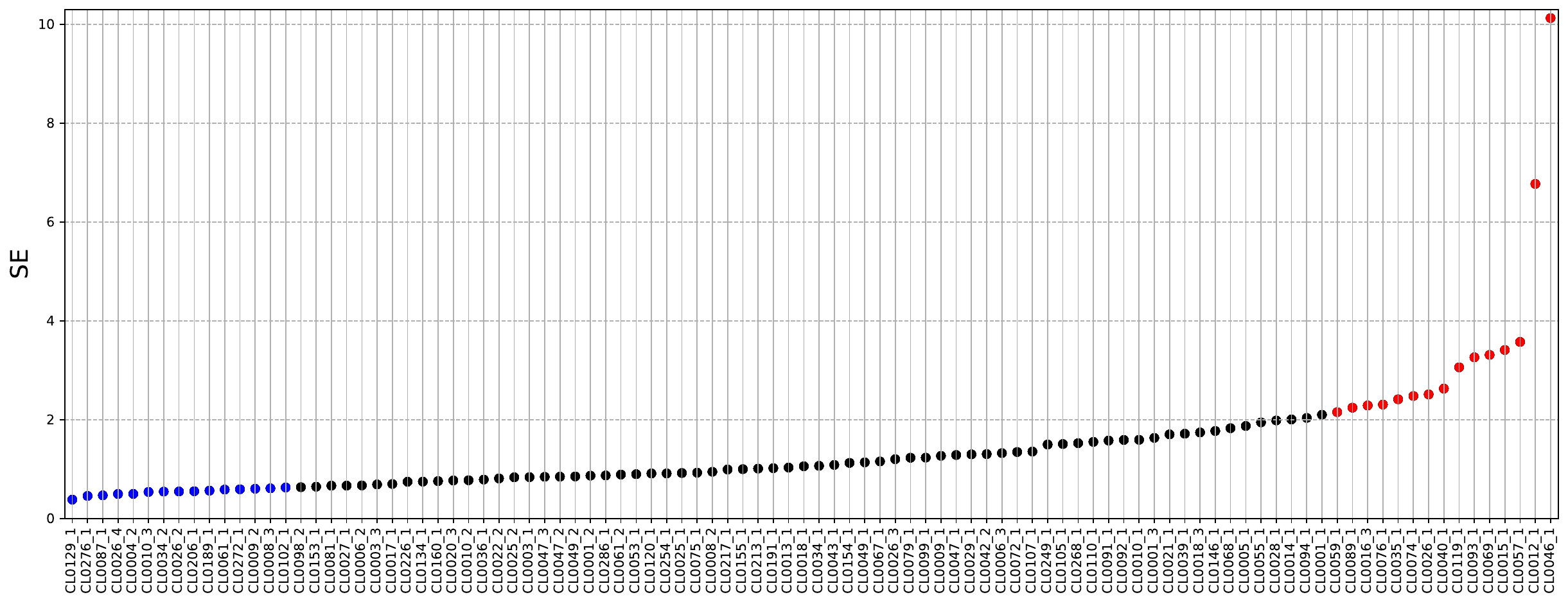}
        \caption{Shape estimator for each cluster in the sample, showed in ascending order, from the most regular to the most irregular. The blue clusters compose the regular sub-sample, the red clusters compose the irregular sub-sample.}
        \label{fig:app_se_ord}
\end{figure*}  

\begin{figure*}[t]  
        \centering
        \includegraphics[width =0.49\textwidth]{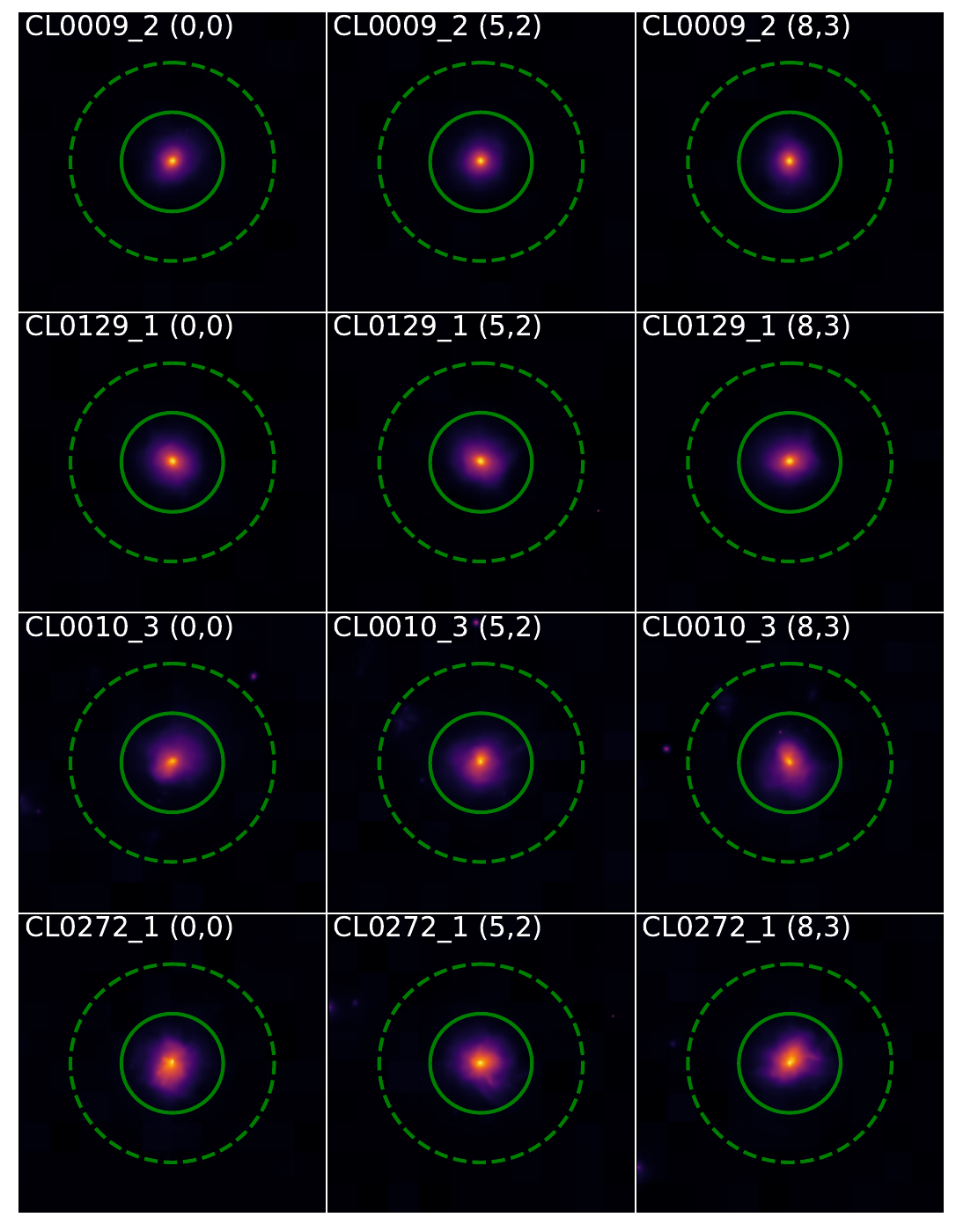}
        \includegraphics[width =0.49\textwidth]{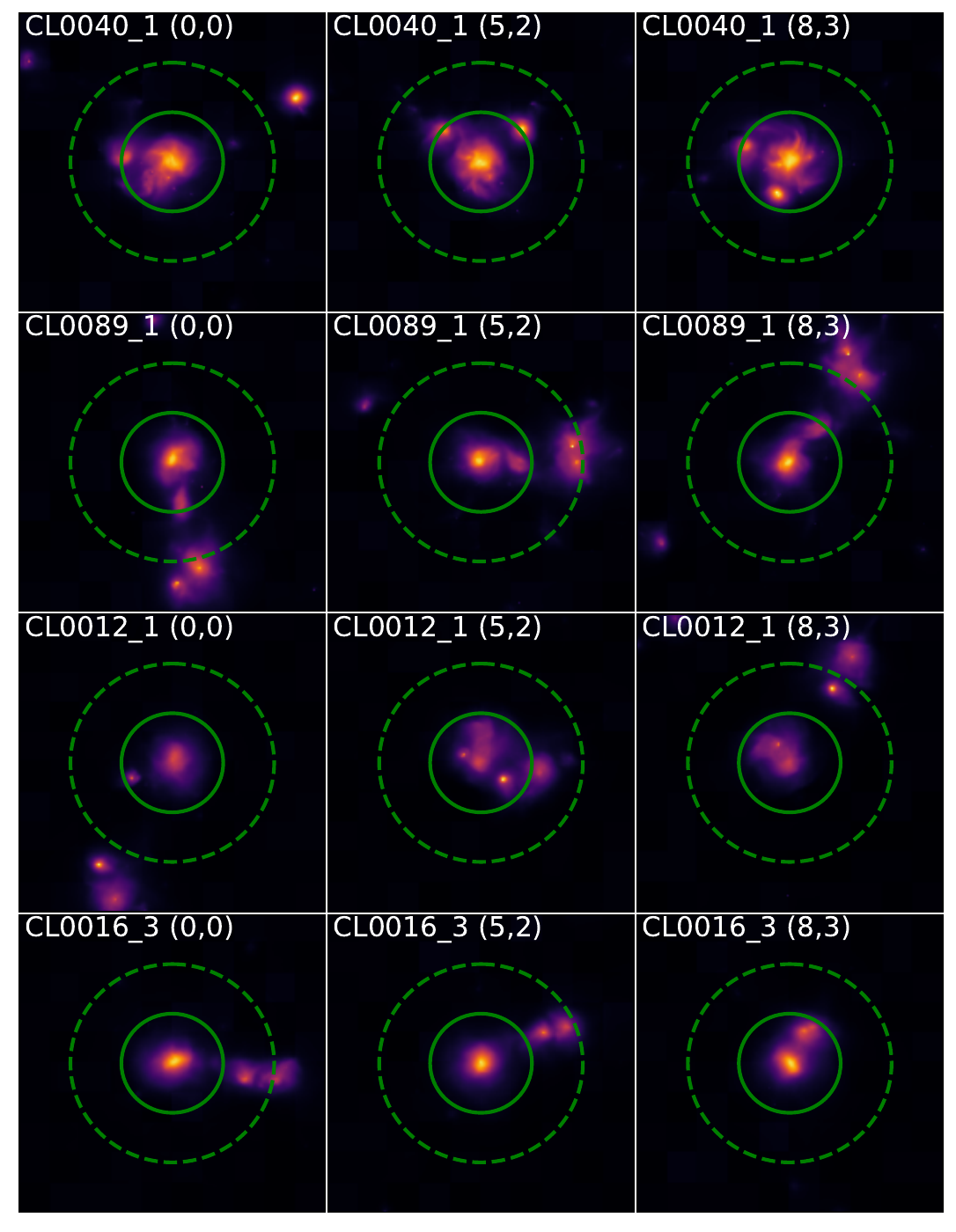}
        \caption{Example of clusters that compose the regular (left) and the irregular (right) sub-sample. For each clusters we report three different lines of sight (i.e. each column is a different sample realization). The solid and dashed circles represent respectively $R_{500}$ and $2 R_{500}$.}
        \label{fig:app_subsample}
\end{figure*}  

We define the \textit{shape estimator}, a quantity that can evaluate the presence of substructures.
For each cluster the \textit{shape estimator} $SE$ is defined as: 
    \begin{equation}
        SE = \sqrt{\sum_i{\sigma^2_{SB}(r_i)}}.
    \end{equation}
where $\sigma_{SB}(r_i)$ is the scatter of the surface brightness profiles (normalized over the mean value) measured from the 40 different lines of sight, evaluated at $R = r_i$.
Substructures produce an alteration in the surface brightness profile by introducing an $SB$ peak where the substructure appears, so that in that region the surface brightness scatter is higher. \IB{The scientific aim of this estimator is to identify the most extreme cases in our sample. Its calibration for a more general use is beyond the scope of this work.}

We report in Figure~\ref{fig:app_se_i} the scatter at different radii for each cluster in the sample. 
The $\sigma_{SB}(r_i)$ distribution clearly becomes broader as the radius increases: this is due to the fact that in the inner regions substructures are not very distinguishable from the central core since their emission is weaker than the central core, so that the 40 surface brightness profiles of each clusters are very similar to each other, thus the scatter is small; if the substructure appear in the outer regions its emission peak is more evident so that the scatter is higher. Moreover, the latter consideration applies only to irregular clusters (i.e. clusters that present substructures), while for regular clusters the surface brightness profile is not affected by emission peak, so that the scatter remains small regardless of the radius. 
For example, if we consider CL0035\_a its scatter goes from $\sigma_{SB}(R = 0.1R_{500}) = 0.05$ to $\sigma_{SB}(R = 2 R_{500}) = 1.9$, while CL0129\_a goes from $\sigma_{SB}(R = 0.1R_{500}) = 0.02$ to $\sigma_{SB}(R = 2 R_{500}) = 0.2$. Therefore we can conclude that CL0129\_a is more regular than CL0035\_a.

In Figure~\ref{fig:app_se_ord} we show the \textit{shape estimator} $SE$ in ascending order, so that the clusters can be considered in "regularity" order. As we can expect, the sample shows a continuous distribution of $SE$, which means that the sample presents a varied morphology composition. We can therefore consider the sample as representative of a common cluster population. 

We chose the first 15 clusters as the regular sub-sample and the last 15 as the irregular sub-sample.
In Figure~\ref{fig:app_subsample} we report four clusters of each sub-sample that show the validity of the method just defined: the clusters of the regular sample do not show any substructure and present a regular central core, while the irregular sample clusters show complex morphology.


    \end{appendix}

\end{document}